\documentclass[aps, showpacs, prb,superscriptaddress,twocolumn,a4paper]{revtex4-1}
\usepackage{graphicx}
\usepackage{amsmath, amsfonts, amssymb, bm}
\usepackage{color}
\usepackage{multirow}
\usepackage[utf8]{inputenc}
\usepackage[english]{babel}
\usepackage{amsmath,array,amssymb,pst-all}
\usepackage{dsfont}
\usepackage{bbold}

\newcommand{\sugb}[1]{\textcolor{blue}{#1}}

\begin{document}

\title{Molecular Structure, Quantum Coherence and Solvent Effects on the  Ultrafast Electron Transport in  BODIPY--C$_{60}$ Derivatives}
\author{Duvalier Madrid-Úsuga}
\altaffiliation{{\tt{duvalier.madrid@correounivalle.edu.co}}}
\address{Centre for Bioinformatics and Photonics (CIBioFi), Calle 13 No.~100-00, Edificio E20 No.~1069, Universidad del Valle, 760032 Cali, Colombia}
\address{Departamento de Física, Universidad del Valle, 760032 Cali, Colombia}
\author{John H. Reina}
\altaffiliation{{\tt{john.reina@correounivalle.edu.co}}}
\address{Centre for Bioinformatics and Photonics (CIBioFi), Calle 13 No.~100-00, Edificio E20 No.~1069, Universidad del Valle, 760032 Cali, Colombia}
\address{Departamento de Física, Universidad del Valle, 760032 Cali, Colombia}


\begin{abstract}

Photo-induced electron transfer in multi-chromophore molecular systems is defined by a critical interplay  between their core units   configuration (donor, molecular bridge and acceptor) and their system-solvent coupling; these lead to energy and charge transport processes that are key in the design of molecular antennas for efficient light harvesting  and organic photovoltaics.  Here, we quantify the ultrafast non-Markovian dissipative dynamics of electron transfer in  D-$\pi$-A molecular photo-systems comprising BODIPY, Zn-porphyrin, Fulleropyrrolidine, and Fulleroisoxazoline. We find that the stabilization energy of the charge transfer states exhibit a significant variation for different polar (Methanol, THF) and nonpolar (Toluene) environments, and determine such sensitivity according to the molecular structure and  the electron-vibration couplings that arise at room temperature. For the considered donor-acceptor ($D$-$A$) dyads we show that the stronger the molecule-solvent coupling the larger  the electron transfer rates, regardless the dyads electronic coherence properties. We found such coupling strengths to be largest  (lowest) for Methanol (Toluene), with an electron transfer rate difference of two orders of magnitude between the polar and nonpolar solvents. For the considered donor-bridge-acceptor ($D$-$B$-$A$) triads, the molecular bridge introduces an intermediate state that allows the realization of $\Lambda$ or cascaded-type energy mechanisms.  We show that the latter configuration, obtained for  {\textit{BDP-ZnP-[PyrC$_{60}$]}} in Methanol, exhibits the highest transfer rate of all the computed triads. Remarkably, and in contrast with the dyads, we show that the larger charge transfer rates are obtained for triads that exhibit  prolonged electron coherence and populations oscillations. 
\end{abstract}

\pacs{03.65.Yz, 
87.15.ht,       
82.20.Xr.       
87.15.ag,       
04.25.-g        
34.70.+e        
}
\maketitle


\section{Introduction}

Dissipation in quantum systems is a phenomenon that plays a prominent role in many problems in physics and chemistry~\sugb{\cite{weiss2006, breuer2007, nitzan2006,weiss2008,reina2002}}. An accurate description of quantum dynamics is thus crucial to the understanding of such phenomena, which include photosynthesis, vibrational energy redistribution, energy transfer, and charge transfer~\sugb{\cite{Scholes2017, collini2009, cheng2009, chenu2015, phelan2019}}. A common approach to this problem is based on coupling a quantum mechanical system with a bath (or environment)~\sugb{\cite{billing2003, Tanimura2006,reina2002,reina2009, forgy2014, mazziotti2012, rebentrost2009}}, where electronic dynamics, vibronic phenomena, as well as quantum coherence play a functional role in various natural and artificial processes~\sugb{\cite{Scholes2017, madrid2019, reina2009, wittmann2020, bredas2017,romero2014, ishizaki2012}}. Intramolecular electronic transfer has been observed in different works in the last decade~\sugb{\cite{collini2009, cheng2009, chenu2015, phelan2019}}. This process plays an important role in many biological, chemical, and physical processes, including photosynthesis, redox chemical reactions, and electron transfer in molecular electronics~\sugb{\cite{fuller2014, song2014, rury2016, rafiq2018}}. Such transfer is often influenced by the dynamics of their molecular or atomic environments, and are accompanied by the dissipation of energy into such environments~\sugb{\cite{berger2020}}. A detailed understanding of such processes is essential for their control and possible exploitation in, e.g., quantum information science~\sugb{\cite{wasielewski2020, atzori2019, reina2018, prando2017, hildner2013,nature2019}}.

A quantum system loses coherence information in its dynamic evolution resulting from the inevitable environmental coupling~\sugb{\cite{breuer2007, weiss2006,reina2002}}. A better understanding of the dynamics of open quantum systems is paramount  to  preventing or controlling quantum  decoherence and the so-called  quantum to classical transition~\sugb{\cite{suter2016, di2016, kang2017, cai2019}}. Experimental results reported during the last decade or so give supporting evidence that electronic quantum coherence aid in the efficient transport of energy and charge from light-harvesting antennas to photosynthetic reaction centres in  biological  systems, even when incoherent natural light is used as a source of excitation~\sugb{\cite{fleming2012,lambert2013, romero2017, powell2017,mirkovic2017}}.

In this work, we address this question for the case of organic molecular complexes, since we are interested in D-$\pi$-A compound for light harvesting tuning and photovoltaic applications. We  investigate the effects due to quantum coherence, electronic structure, and vibronic phenomena on the electron dynamics that take place in organic molecules comprising donor-bridge-acceptor units: BODIPY, Zn-porphyrin, and Pyrrolidine (and Isoxazoline) rings. By changing the molecular structure associated to such compounds and the vibrational features due to the different  solvents here considered (Methanol, THF and Toluene), we show how to tailor the electronic dynamics and charge transfer rates. This study is not only useful for our understanding of electron  transfer processes, but also relevant to the advancement of, e.g. organic photovoltaics based on coherence effects in photosynthetic systems and supra-molecular architectures~\sugb{\cite{Scholes2017,jackson2015,wittmann2020,bredas2017,romero2014, ishizaki2012, sengupta2013, falke2014, ryno2016}}, and emergent molecular quantum technologies~\sugb{\cite{wasielewski2020,nature2019,atzori2019,reina2018,avinash2018,prando2017,hildner2013,reina2017}}.

The considered molecular systems derived from BODIPY are coupled to a solvent described as a bath of bosonic modes in order to model the strong electron–vibration couplings, with corresponding spectral density features obtained for each BODIPY derivative: the site energies and electronic couplings are calculated by using a continuous polarizable model for different solvent environments, and site-to-site couplings are computed by means of the generalized Mulliken Hush method at the DFT level. The molecular systems are modelled as two- and three-site systems, and we use the density functional theory (DFT) for obtaining the site energies required in the construction of the open system diabatic Hamiltonians relevant to the computation of coherence and electron transfer properties.

\subsection{Molecular Systems} \label{Model_Systems}

\begin{figure*}[htp]
\centering
\includegraphics[scale=0.8]{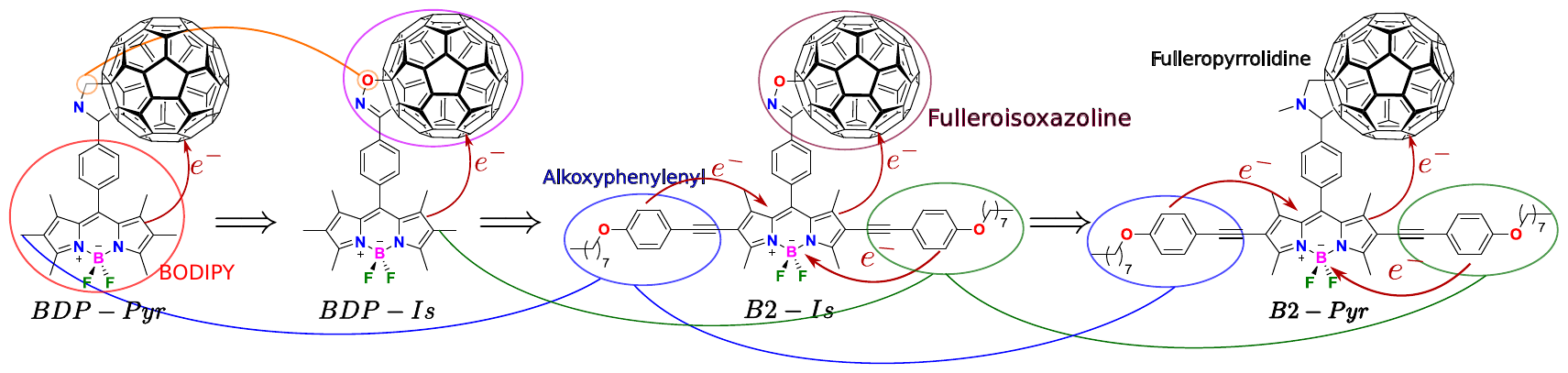}
\caption{Molecular structure of the chromophores under study. The molecular systems are formed from BODIPY and BODIPY+alkoxyphenylethynyl as donors and Fulleropyrrolidine and Fulleroisoxazoline as acceptors. The considered solvents, Methanol, Toluene and Tetrahydrofuran (THF),  have corresponding dielectric constants $\epsilon_{_M} = 32.613$, $\epsilon_{_H}=7.4257$, and  $\epsilon_{_T} = 2.374$.}
\label{Fig1F}
\end{figure*}

 We consider D-$\pi$-A compounds derived from BODIPY. We have chosen, for the sake of analysis, to specialize to the molecular systems \textit{BDP-Pyr}, \textit{BDP-Is}, \textit{B2-Pyr}, and \textit{B2-Is}, since they exhibit a favorable electronic transport~\sugb{\cite{cabrera2017, cabrera2018,calderon2021,madrid2018}}. Such structures are constructed from \textit{Borrondipyrromethenes Fulleropyrrolidine} for \textbf{\textit{BDP-Pyr}}, \textit{Borrondipyrromethenes Fulleroisoxazoline} for \textbf{\textit {BDP-Is}}, 2,6-di[4'-(n-octyloxy)phenylethynyl] -4,4-Difluoro-8-phenyl-1,3,5,7-tetra- methyl -4-bora-3a, 4-adiaza-s-indacene Fulleropyrrolidine for \textbf{\textit{B2-Pyr}}, and 2,6-di[4'-(n-octy- loxy)phenylethynyl] -4,4-Difluoro-8-phenyl-1,3,5,7-tetramethyl -4-bora-3a,4-adiaza-s-indacene Fulleroisoxazoline for \textbf{\textit{B2-Is}}, as schematically shown in Fig.~\sugb{\ref{Fig1F}}. In what follows, we shall analyze these systems and their molecular structure effects on the process of electronic transfer, quantum coherence, and molecular systems-solvent coupling. 

The considered \textbf{\textit{BDP-Pyr}} and \textbf{\textit{BDP-Is}} systems allow to highlight the differences when considering the inclusion of the heterocyclic rings of Pyrrolidine and Isoxazoline. Substitution of a carbon atom for an oxygen atom to generate the Isoxazoline ring is expected to increase the inductive effect of the ring and thus increase the capacity of the functionalized fullerene to accept electrons, further favoring electronic interactions between BODIPY and the \textbf{C}$_{60}$.
The Isoxazoline ring is regarded as a good promoter of electronic injection if used as a bridge between \textbf{C}$_{60}$ and BODIPY, and  theoretical evidence of such electron-attracting effect of the Isoxazoline ring on fullerene has been reported in~\sugb{\cite{Langa2000}}: both  $sp^3$ atoms of the  \textbf{C}$_{60}$ cage support a higher positive charge density in the Fulleroisoxazoline derivative than in the analogous Fulleropyrrolidine, suggesting that the former has a better electron affinity as a result of the electronegativity of the oxygen atom and the electron-deficient character of the carbon $sp^2$ directly attached to \textbf{C}$_{60}$.

On the other hand, the design of architectures with multiple units that help capture external radiation and therefore extend the absorption range has been an attractive strategy to improve the efficiency of artificial photo-synthetic systems. In this sense, we consider the \textbf{\textit{B2-Is}} and \textbf{\textit{B2-Pyr}} systems based on the covalent bonding of alkoxyphenylethynyl groups in the molecular structure of BODIPY, which help to increase the absorption spectral range, being of particular interest to favor charge transfer processes.

\subsection{Molecular plus Vibronic Coupling Hamiltonian}

For the analysis here presented, we start from a molecular system's Hamiltonian $\widehat{H}_S$ that can be written as:
\begin{equation}
\widehat{H}_S=\sum_j\varepsilon_j\vert j\rangle\langle j\vert + \sum_{j\neq i} V_{ji}\vert j\rangle\langle i\vert,
\label{Ecu1F}
\end{equation}
where the state $\vert j \rangle$ describes a scenario where the charge is spatially located on the $j$'th site, $\varepsilon_j$ is the energy state \textit{j}, and $ V_ {ji} $ is the electronic coupling parameter connecting the states \textit{j} and \textit{i}. To construct the Hamiltonian for the molecular systems we need to calculate the diabatic electronic couplings, $\lbrace V_ {ji} \rbrace $, and the energies, $\lbrace \varepsilon_j \rbrace $. The challenge posed here is how to transform adiabatic states from quantum chemical calculations to diabatic states. In this work, we use the generalized Mulliken-Hush (GMH) theory~\sugb{\cite{cave1996, zheng2017}} to obtain these couplings.
\medskip

\begin{figure}[htp]
\centering
\includegraphics[scale=0.4]{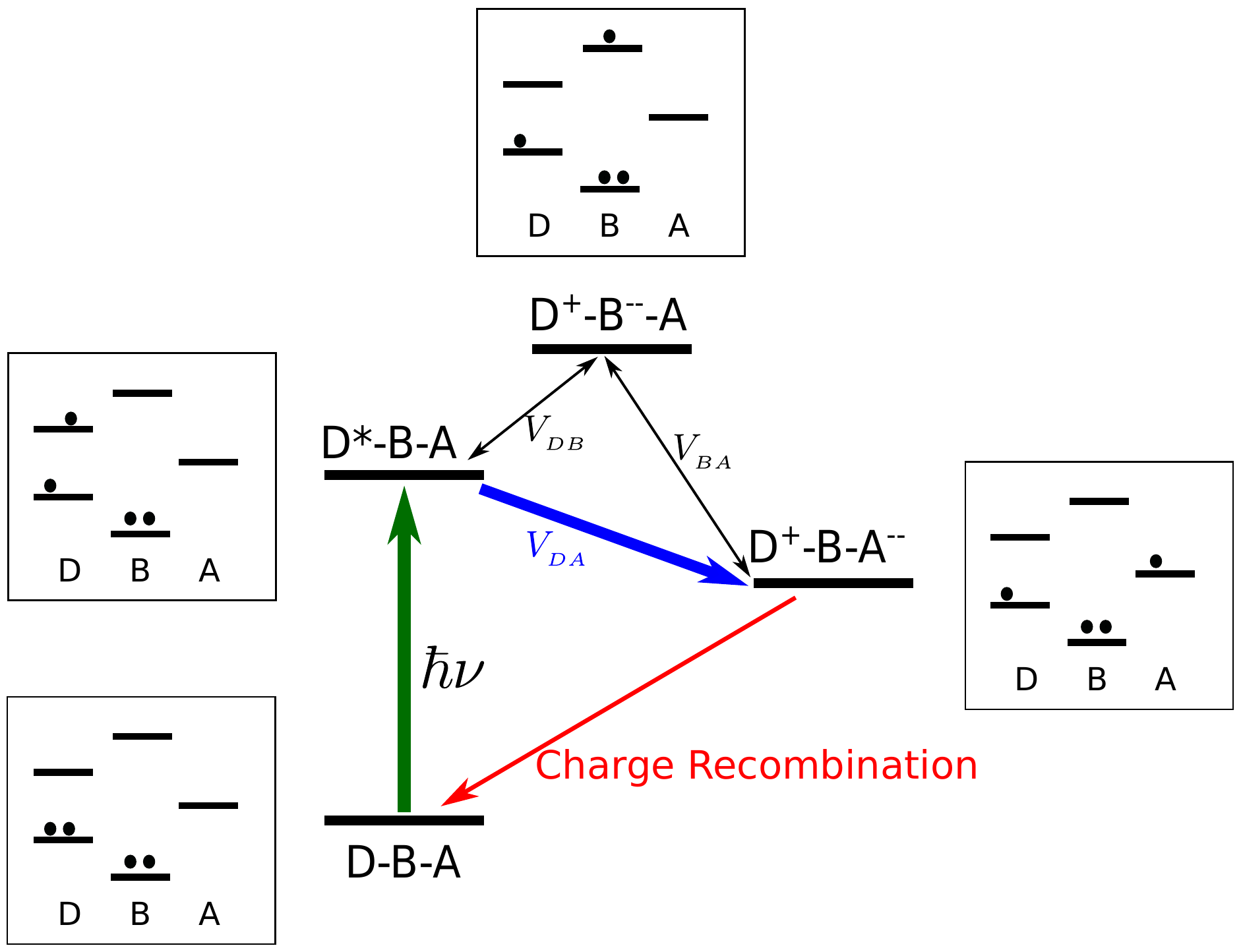} 
\caption{Jablonski schematic diagram illustrating the relevant states for $\widehat{H}_s$. Initially, the system is in the electronic basis ($D-B-A$) state. Once the electron-donor fragment has been photo-excited, the charge transfer reaction is possible, and the creation of a charge transfer state can take place. The absorption of light by BODIPY initiates the electron transfer process. This is followed by an electron transfer from $D$ to $B$ to reach the state of charge Transfer ($D^+- B^--A$), and finally the transfer to the electron-acceptor fragment to reach the target state of charge separation ($D^+-B-A^-$). The various states are sketched in terms of electronic configurations; see boxes. Couplings are represented by double arrows connecting the states. In addition to these direct electron transfer reactions, the system can recombine by charging from the partial or total charge transfer state (red arrows).}
\label{Fig2F}
\end{figure}

In order to choose adiabatic states to be used in the construction of diabatic ones, we require, in addition to the ground state, an appropriate identification of  excited states that best represent the desired diabatic states  (see Fig.~\sugb{\ref{Fig2F}}). First, we implement Density Functional Theory (DFT) with the interchange-correlation functional CAM-B3LYP (Becker three-parameter Lee-Yang-Parr)~\sugb{\cite{becke1993, ganji2015, ganji2016}} and the basis set  6-31G(d, p), by using Gaussian 09~\sugb{\cite{gaussian09}} to investigate the optimization geometry of the D-$\pi$-A Compounds. From the optimized ground state geometry, the lowest 30 vertical excitation energies were calculated using time-dependent (TD)-DFT. The effects due to the solvent coupling to the molecular systems were modelled within the framework of a continuous dielectric medium for the optimization of the molecular systems in the basis and excited states, following a Conductor-like Polarizable Continuum Model (C-PCM)~\sugb{\cite{takano2005, chiu2016}}. The modelled  solvents, Methanol, Tetrahydrofuran (THF), and  Toluene,  have, respectively,  dielectric constants $\epsilon_{_M} = 32.613$, $\epsilon_{_H}=7.4257$, and $\epsilon_{_T} = 2.374$. The choice of such solvents obey to their moderate dielectric constants and to an ease in the solubilization of  the studied systems.

\begin{figure}[htp]
\centering
\includegraphics[scale=0.05]{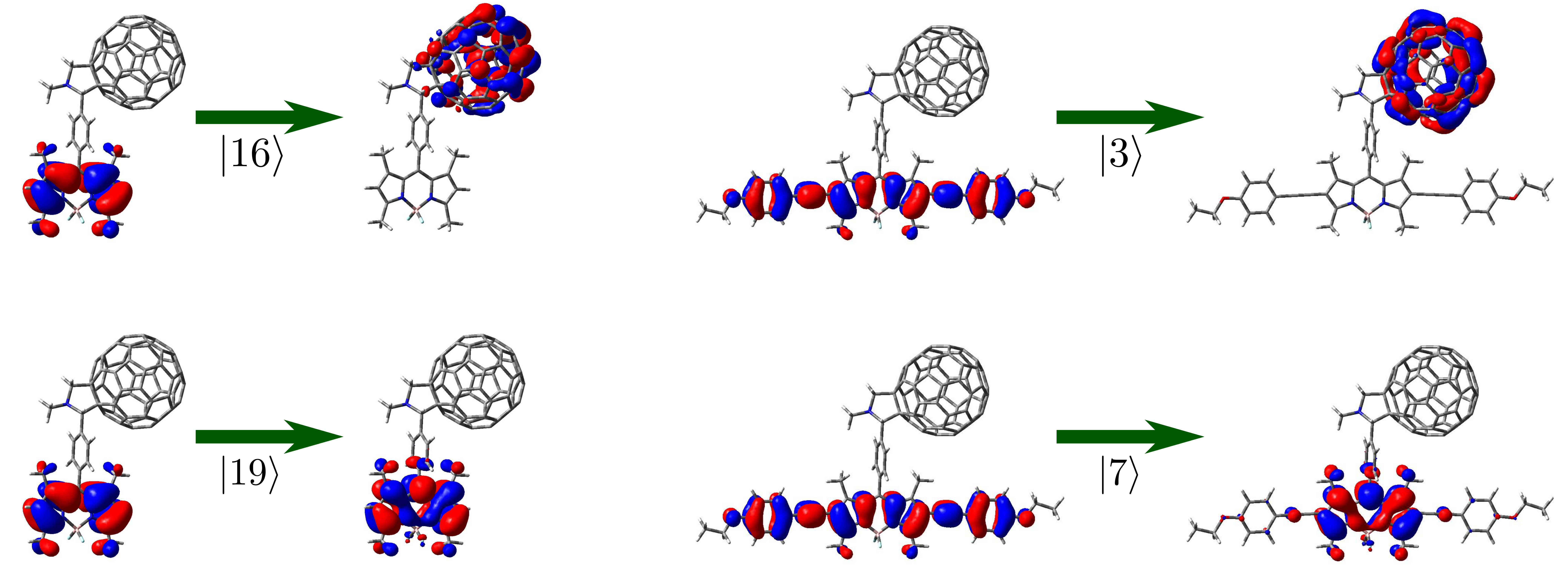} 
\caption{The natural transition orbitals--NTOs are used to identify the relevant adiabatic basis used in the treatment of the molecules' GMH. From the NTOs, the density change is used to find adiabatic states that are more similar to the diabatic state model. The green arrows show the direction of the transition from the NTO ``HOMO'' to the NTO ``LUMO". Kets are used to indicate the adiabatic states, and $\vert N \rangle$ denotes the excited state $N$. In this case they are shown for the  \textbf{\textit{BDP-Pyr}} (left) and \textbf{\textit{B2-Pyr}} (right) systems.}
\label{Fig3F}
\end{figure}

From the TD-DFT  calculations in the optimized ground state geometries, the natural transition orbitals (NTO)~\sugb{\cite{martin2003}} of the 30 excitations were calculated. NTOs help us to obtain a qualitative impression of the location of the electronic transition. Thus, instead of the large number of combinations of orbitals occasionally found in standard TD-DFT calculations, the orbitals combine to provide the most representative orbital pair of electron transfer (see Fig.~\sugb{\ref{Fig3F}}). From the dyads NTOs, the adiabatic states located in the donor and the acceptor are identified; these states are used to create the GMH diabatic states, and to mainly calculate the electronic couplings $ V_{ij} $. It is clear from Fig.~\sugb{\ref{Fig3F}} that the adiabatic states $\vert 3 \rangle$ and $\vert 7 \rangle$ closely model the charge movement associated with two diabatic states of interest for \textbf{\textit{B2-Pyr}}. However, a close similarity of the desired states for \textbf{\textit{BDP-Pyr}} is not easily found among the low-energy adiabatic states, and the closest similarity is found with the $\vert 16 \rangle$ and $\vert 19 \rangle$ states.
\medskip

\begin{figure}[htp]
\centering
\includegraphics[scale=0.35]{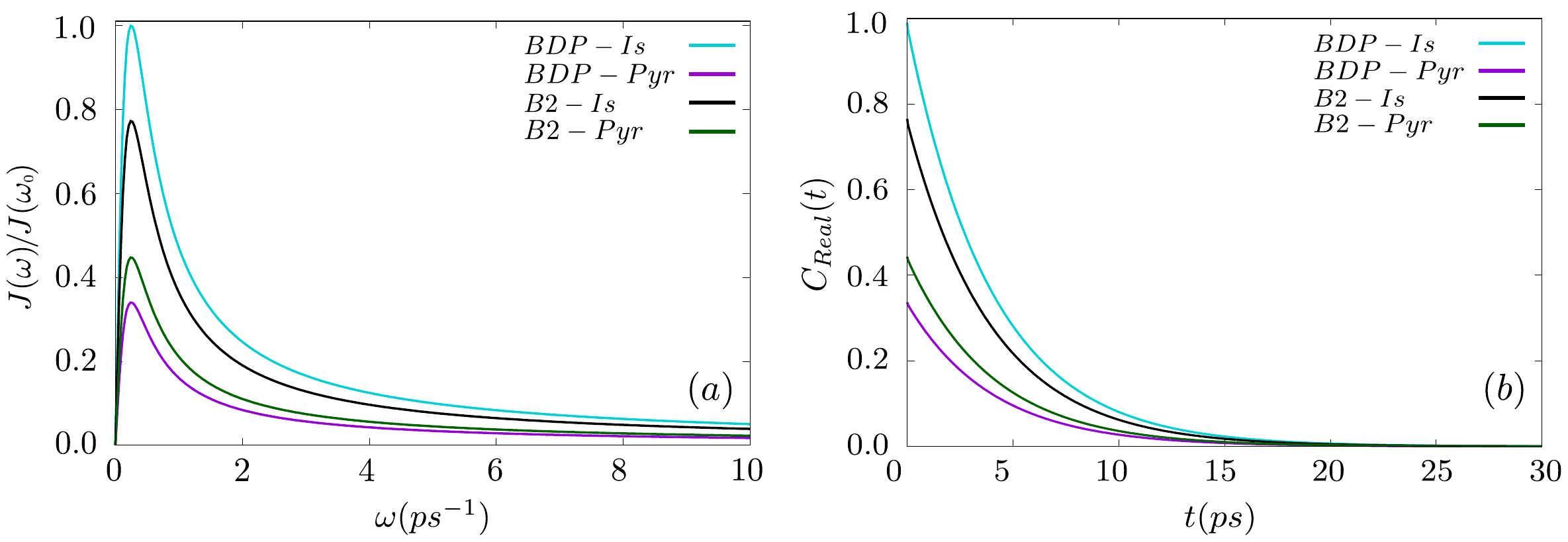}
\caption{(a) The ohmic spectral density for the Drude spectral distribution as a function of the frequency $\omega$, with $\lambda=17$~cm$^{-1}$ (\textbf{\textit{BDP-Is}}), $\lambda=22$~cm$^{-1}$  (\textbf{\textit{B2-Is}}), $\lambda=38$~cm$^{-1}$ (\textbf{\textit{B2-Pyr}}), and $\lambda=50$~cm$^{-1}$ (\textbf{\textit{BDP-Pyr}}), (b) Correlation function $C_{Real}(t)$ plotted for the molecular systems under study, in Methanol, $\omega_c=0.250$~ps$^{-1}$, and  $T=300$~K.}
\label{Fig4F}
\end{figure}

As for the diabatic state energies used in the construction of $\widehat{H}_S$, they are calculated with a combination of excitation energies from TD-DFT computations, by  implementing the ionization potential ($IP$) and the electronic affinity ($EA$) of the different charge sites, and a Coulomb energy term that describes the interaction of the relevant charge sites. The ground state is the reference state and its diabatic state energy is set to zero. The energy of the locally excited diabatic state ($D^{*}$--$ A$) is calculated as the vertical excitation energy to the adiabatic state that most represents it ($\vert 19 \rangle$). The energy of one of the charge transfer states (CTS) with a positive charge at site X and a negative charge at site Y is calculated as~\sugb{\cite{blomberg1998, powell2017, storm2019}}:
\begin{equation}
\varepsilon_j=IP_{_D}-EA_y+E_c(r_{xy}),
\end{equation}
where $IP_D$ is the ionization potential of the  donor fragment, which was calculated by subtracting the total energy of the
neutral molecule from that of the cation, as computed by DFT. $EA_y$ is the electronic affinity of the acceptor fragment  and was obtained by subtracting the total energy of the anion from that of the neutral state. $E_c(r_{xy})\sim  1/\varepsilon_s r_{_{xy}}$ is the Coulomb interaction between the cation on the  donor fragment and the anion on the  acceptor fragment, and  $\varepsilon_s$ denotes the solvent dielectric constant. $r_{Dy}$ is the distance from the center of mass of the  donor fragment to the center of mass of the acceptor fragment,  according to the optimized ground state structure.

We treat the molecular system ($\widehat{H}_S$) as an open quantum system, and hence each considered site is coupled to a phonon bath ($B$) comprising the vibrations of the solvent' molecule, which in turn is characterized by an infinite number of oscillators, and its Hamiltonian can be written as
\begin{equation}
\widehat{H}_{_B}=\sum_{\xi}\Bigg(\frac{\widehat{p}_{_{\xi}}^2}{2m_{_{\xi}}}+\frac{m_{_{\xi}}\omega_{_{\xi}}^2\widehat{x}_{_{\xi}}^2}{2}\Bigg),
\end{equation}
where \textit{$\xi$} indexes the bath modes, with coordinates $\widehat{x}_{_{\xi}}$, moments $\widehat{p}_{_{\xi}}$, and masses $m_{_{\xi}}$. The vibronic interaction between each site and its bath defines the system-bath coupling Hamiltonian, $\widehat{H}_{_{SB}}$, given by
\begin{equation}
\widehat{H}_{_{SB}}=-\sum_{_{j,\xi}}g_{_{j,\xi}}\widehat{\sigma}_z\widehat{x}_{_{\xi}},
\end{equation}
where $\widehat{\sigma}_z$ represents the system operator, $\widehat{x}_{_{\xi}}$ is the bath coordinates operator, and  $g_{_{j,\xi}}$ gives the system-bath coupling strength.

The bath effects are captured through the quantum bath correlation function, which can be calculated as:
\begin{equation}
C(t)=\frac{1}{\pi}\int_0^\infty J(\omega)[\coth(\beta\omega/2)\cos(\omega t)-\textit{i}\sin(\omega t)]d\omega,
\label{EcuFF}
\end{equation}
where $\beta\equiv 1/k_BT$, with temperature $T$ and Boltzmann constant $k_B$. We consider a spectral density function $J(\omega)$ of the Drude-Lorentz type~\sugb{\cite{tanimura2015,tanimura2014,tanimura2020,JGilmore, Gilmore,madrid2019,reina2009}}
\begin{equation}
J(\omega)=\frac{2\lambda\omega_c\omega}{\omega^2 + \omega_c^2},
\label{Drude}
\end{equation}
where the cutoff $\omega_c$ represents the width of the spectral distribution of the collective bath modes and is the reciprocal of the relaxation time due to the solvent, $\tau_s = 1/\omega_c $. The reorganization energy $\lambda$ influences the function's amplitude and the system-environment coupling $\eta=\frac{\lambda}{2h\omega_c}$. 

By using Eq. (\sugb{\ref{Drude}}), we get $C(t)=L_R(t)-iL_I(t)$, where the relaxation function is temperature-independent and is expressed as $L_I(t)=\widetilde{c}_{_0}e^{-\nu_0\vert t\vert}$, with $\widetilde{c}_{_0}=4\pi\hbar\omega_c^2$, while the noise correlation is given by $L_R(t)=c_{_0}e^{-\nu_{_0}\vert t\vert}+\sum_k c_{_k}e^{-\nu_{_k}\vert t\vert}$, with $c_{_0}=4\pi\hbar\omega_c^2\cot(\beta\hbar\omega_c)/2$ and $c_{_k}=16\pi\omega_c^2/\beta(\nu^2-\omega_c^2)$ for the $k$th Matsubara frequency $\nu_{_k}=2\pi k/\beta\hbar$, and $\nu_{_0}=\omega_c$.\sugb{\cite{Tanimura2006,tanimura1989,tanimura1990,ishizaki2005}}. The correlation function is made up of two types of non-Markovian components: one is of mechanical origin and is characterized by dissipation ($L_I(t)$) and fluctuations ($L_R(t)$) expressed in terms of $e^{-\omega_c t} $, and the other is of quantum thermal origin and is characterized by fluctuations only ($L_R(t)$). When $\omega_c$ is much larger than $\nu_1$, the mechanical contribution with $e^{-\omega_2 t}$ vanishes for $t>1/\nu_1$, and then, the effects due to quantum thermal noise arise. Figure~\sugb{\ref{Fig4F}} shows the spectral density for the studied systems in Methanol, as well as the real part of the bath correlation function at $T=300$~K;  the $L_R(t)$ profile  exhibits a non-Markovian nature for the considered systems.

\section{Results and Discussion}

We begin by considering a dissipative two-site system, where the system Hamiltonian of Eq.~\sugb{\ref{Ecu1F}} has site energies $\varepsilon_1$, $\varepsilon_2$ and coupling $V_{ij}=V_{DA}$. Each site is coupled independently to its own thermal bath, with a Drude-Lorentz spectral density given by Eq. (\sugb{\ref{Drude}}). Since, in principle, each site is connected to its own thermal bath, the spectral densities could vary from site to site; for simplicity, we first consider that the system-bath coupling is the same for both sites. The simulation parameters were chosen for the corresponding $\tau_s$ and $V_{DA}$ values associated with each molecular system, in three different solvents, at room temperature $T=300$~K (see Table~\sugb{\ref{Tab1F}}). 

\begin{table*}[htp]
\caption{Relevant parameters of the $D-A$ system plus solvent dynamics. Energy difference between donor and acceptor states $\Delta\varepsilon=\vert E_D-E_A \vert$, electronic coupling $V_{DA}$, and system-solvent coupling $\eta=\frac{\lambda}{2h\omega_c}$.}
\begin{center}
\scalebox{0.6}{
\begin{tabular}{lcccccccccccc}
\hline
\hline
\\
 & & \Large{\textbf{\textit{BDP-Is}}} & & & \Large{\textbf{\textit{BDP-Pyr}}} & & & \Large{\textbf{\textit{B2-Is}}} & & & \Large{\textbf{\textit{B2-Pyr}}}\\
\multirow{2}{*}{\Large{\textbf{Solvent}}} &&&&&&&&&&&\\
\cline{2-13}
\\
 & \large{\textbf{$\Delta\varepsilon$ (cm$^{-1}$)}} & \large{\textbf{\textit{V}$_{_{DA}}$ (cm$^{-1}$)}} & \Large{\textbf{$\eta_{_1}$}} & \large{\textbf{$\Delta\varepsilon$ (cm$^{-1}$)}} & \large{\textbf{\textit{V}$_{_{DA}}$ (cm$^{-1}$)}} & \Large{\textbf{$\eta_{_2}$}} & \large{\textbf{$\Delta\varepsilon$ (cm$^{-1}$)}} & \large{\textbf{\textit{V}$_{_{DA}}$ (cm$^{-1}$)}} & \Large{\textbf{$\eta_3$}} & \large{\textbf{$\Delta\varepsilon$ (cm$^{-1}$)}} & \large{\textbf{\textit{V}$_{_{DA}}$ (cm$^{-1}$)}} & \Large{\textbf{$\eta_4$}} \\
\\
\hline
\hline
\\
\large{\textbf{Methanol}}         & $2564$ & $204$ & $1.0193$ & $1975$ & $153$ & $3.0950$ & $1604$ & $100$ & $1.3645$ & $894$  & $75$  & $2.3265$ \\
\large{\textbf{Tetrahidrofurano}} & $2285$ & $280$ & $0.1573$ & $2095$ & $210$ & $0.7896$ & $2019$ & $160$ & $0.2021$ & $1230$ & $135$ & $0.3715$ \\
\large{\textbf{Toluene}}          & $2340$ & $388$ & $0.0158$ & $2507$ & $300$ & $0.0316$ & $570$  &  $62$ & $0.0178$ & $2134$ & $170$  & $0.0278$ \\
\\
\hline
\hline
\end{tabular}}
\end{center}   
\label{Tab1F}
\end{table*}

Depending on the $V_{ij}/\omega_c$ ratio between the electronic coupling and the  characteristic bath frequency, electron transfer reactions are commonly classified as adiabatic or nonadiabatic. In the adiabatic reactions, the characteristic timescale of the bath is slow compared to the electronic tunneling time, hence $V_{ij}/\omega_c>1$. In nonadiabatic reactions, $\tau_s$ is fast compared to the time scale of electron tunneling, and $V_{ij}/\omega_c<1$. While in many solid-state physics applications of the spin-boson model, the nonadiabatic regime is of  primarily interest, in chemical physics, and in particular in the context of electron-transfer reactions, both the adiabatic and intermediate regime have many applications.

\begin{figure}[ht]
\centering
\includegraphics[scale=0.35]{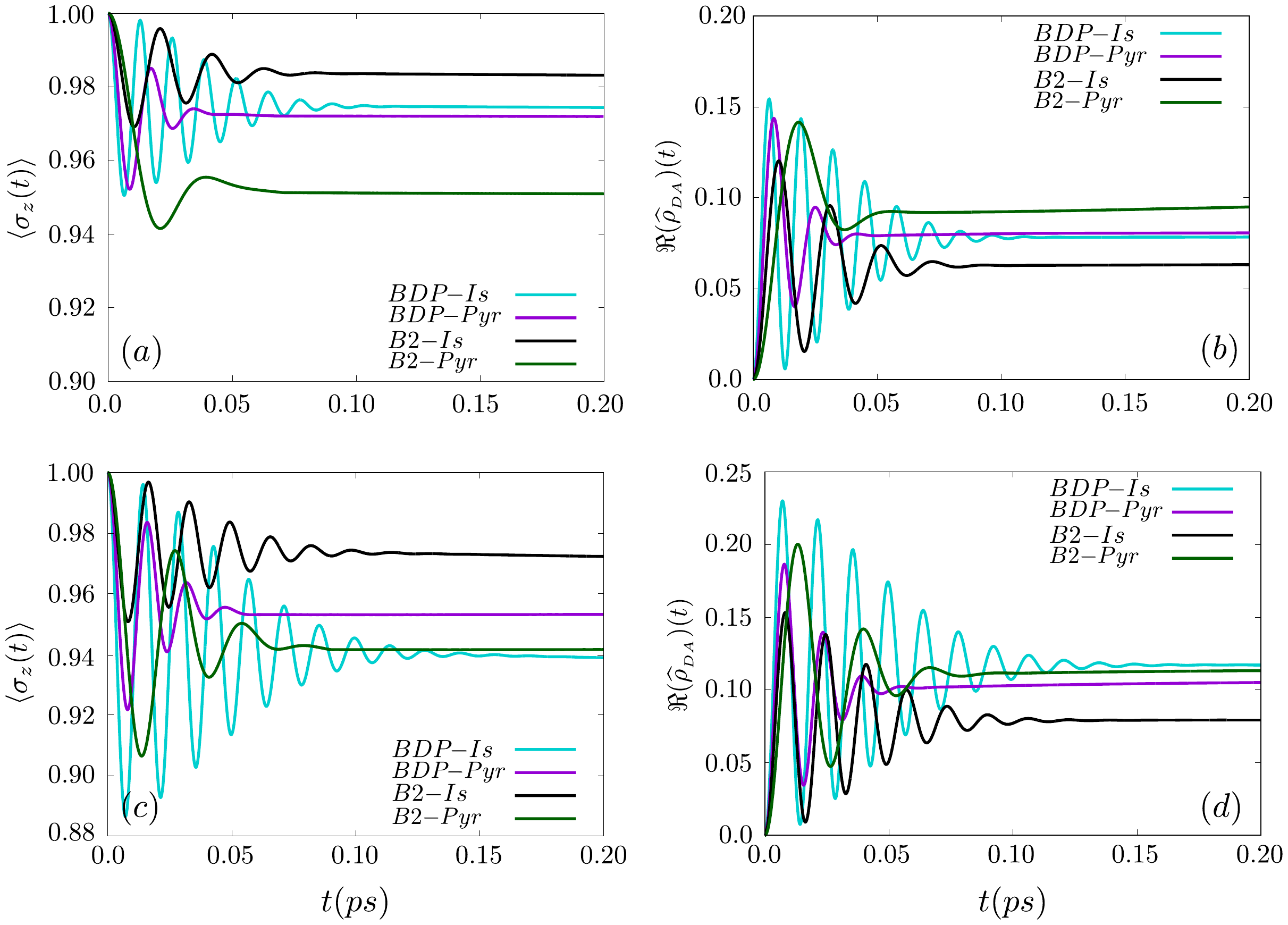}
\caption{(a)~Population inversion $\langle\widehat{\sigma}_z(t)\rangle$ and (b) Real part of the quantum coherence $\rho_{DA}$, for: \textbf{\textit{BDP-Is}} ($\eta_{_1}=1.0193$, turquoise curve), \textbf{\textit{BDP-Pyr}} ($\eta_{_2}=3.0950$, purple curve), \textbf{\textit{B2-Is}} ($\eta_{_3}=1.3645$, black curve) and \textbf{\textit{B2-Pyr}} ($\eta_{_4}=2.3265$, green curve)in Methanol. (c)~Population inversion $ \langle \widehat{\sigma}_z(t) \rangle $, and (d) Real part of the quantum coherence $\rho_{DA}$, for: \textbf{\textit{BDP-Is}} ($\eta_{_1}=0.1573$, turquoise curve), \textbf{\textit{BDP-Pyr}} ($\eta_{_2}=0.7896$, purple curve), \textbf{\textit{B2-Is}} ($\eta_3=0.2061$, black curve) and \textbf{\textit{B2-Pyr}} ($\eta_{_4}=0.3715$, green curve), in Tetrahidrofurano (THF) with $\langle\sigma_z(0)\rangle=1$ at $T=300$~K.}
\label{Fig5F}
\end{figure}

The results for the two sites electronic dynamics are shown in Fig.~\sugb{\ref{Fig5F}}, demonstrating the transfer of electrons from the photoexcited donor state to the acceptor state. These calculations involve a simple donor-acceptor transfer process and therefore there is no possibility of interferometric behavior because there is only one available route. The damped oscillations present in the graphs arise from the nature of electron transfer, since there is a sufficiently strong donor-acceptor electronic coupling in relation to the energy gap between them. In addition, we note that the system's dynamics is of the adiabatic type since for all the molecules, $V_{DA}/\omega_c>1$. The electronic coupling for \textbf{\textit{B2-Pyr}} has been found to be smaller in the presence of Methanol and THF than in Toluene, while in the presence of the latter, the \textbf{\textit{B2-Is}} exhibit the lower coupling (see Table~\sugb{\ref{Tab1F}}); this is probably due to a poor overlap resulting from the almost perpendicular geometry of BODIPY relative to the Fulleropyrrolidine and the  Fulleroisozaxoline linker.

 Table~\sugb{\ref{Tab1F}} shows that the substitution of the ring of pyrrolidine by one of isoxazoline increases the  energy gap $\Delta\varepsilon$  of the systems, as we see when comparing \textbf{\textit{BDP-Is}} and \textbf{\textit{BDP-Pyr}}, in  Methanol and THF; on the other hand, if such process is done in Toluene, such energy gap decreases. Furthermore,  the inclusion of the alkoxyphenylethynyl group further obeys the same trend, but with reduced $\Delta\varepsilon$  values for \textbf{\textit{B2-Is}} and \textbf{\textit{B2-Pyr}}, which will be reflected in a higher probability of charge transfer in these latter systems, since the energy levels are more closely spaced, which favors the driving force. 

On the other hand, the polarity of the solvent shows a considerable effect on the properties of the molecular systems, and in particular on the energy gap $\Delta\varepsilon$ and the electronic coupling $V_{DA}$. For  \textbf{\textit{BDP-Pyr}} and \textbf{\textit{B2-Pyr}}, such values increase as the solvents' dielectric constants decrease  ($\epsilon_{_M}> \epsilon_{_H}>  \epsilon_{_T}$). However, when looking at the \textbf{\textit{BDP-Is}} and \textbf{\textit{B2-Is}} systems,  $\Delta\varepsilon$ and $V_{DA}$ do not exhibit a well-defined trend as in the previous case. For example, for the system \textbf{\textit{BDP-Is}} the value of $\Delta\varepsilon$ decreases when passing from the solvent Methanol to THF; however, this value increases when considering the interaction with Toluene. In the case of  \textbf{\textit{B2-Is}}, $\Delta\varepsilon$ increases when decreasing the dielectric constant (Methanol to THF), but its value decreases when considering the interaction with Toluene (which has the lowest dielectric constant of all the considered solvents). Another point to consider here (see Table~\sugb{\ref{Tab1F}}) is the strength of the system-solvent interaction, $\eta$, which falls within the strong coupling regime, where a non-Markovian dynamics description is required. These results indicate that the molecular configurations here discussed under the effects of polar (Methanol and THF) and non-polar (Toluene) solvents induce a complex dynamics whose coherence and electron transport are non-trivial and require a rigorous treatment. We use a non-perturbative open quantum system approach, with the numerically exact HEOM method in order to compute  the ultrafast dynamics, quantum observables and the electron transfer rates of the described molecular systems.

Figures~\sugb{\ref{Fig5F}~(a)} and \sugb{\ref{Fig5F}~(c)} display the population inversion $\langle \widehat{\sigma}_z(t) \rangle = \rho_D(t)-\rho_A(t)$ of the electronic transitions as a function of time for molecular systems in Methanol and THF, respectively. In the figures, the populations   $\rho_D(t)$ and $\rho_A(t)$ are considered to be localized on the donor at $t =0$, i.e., $\langle \widehat{\sigma}_z(0)\rangle= 1$. Figures~\sugb{\ref{Fig5F}~(b)} and \sugb{\ref{Fig5F}~(d)} plot the real-part coherence $\Re{(\rho_{DA}(t))}$ of the molecules, where $\rho_{DA}(t)=\rho_{AD}^*(t)$ are the coherences (matrix elements) of the density operator. In order to emulate ambient conditions, all our calculations are performed at room temperature, the most critical scenario for the survival of quantum coherence.

In Fig.~\sugb{\ref{Fig5F}}, the oscillations exhibited by the population inversion of the effective two-level  system are more rapidly damped  in the molecular systems that have a pyrrolidine ring (\textbf{\textit{BDP-Pyr}} and \textbf{\textit{B2-Pyr}}) in their structure ($\sim 40-55$~fs), than those containing an isoxazoline ring ($\sim 75-100$~fs), indicating a stronger coupling between these dyads and their environment, and the consequent  amplitude loss in the populations oscillations. If compared with the THF solvent scenario, as shown in Fig.~\sugb{\ref{Fig5F}~(c)}, the population inversion decays are prolonged  to about 100 fs and 160 fs, respectively. We also note that the inclusion of the alkoxyphenylethynyl group favors the decrease of the oscillations in the population inversion, since they considerably increase the coupling to the environment. The way in which the bath oscillation modes affect the molecular quantum dynamics can also be seen in Figs.~\sugb{\ref{Fig5F}~(a)} and \sugb{\ref{Fig5F}~(c)}, where $\omega_c \sim V_{DA} $ indicates a ``slow'' bath scenario. It is clear that a rapid bath ($\omega_c \gg V_{DA}$) favors neither the amplitude nor the duration time of the populations oscillations, compared to the case $\omega_c\sim V_{DA}$, where the oscillations last for a much longer time as well as their amplitude. In our case, the molecules interact with Methanol and THF ($V_{DA}/\omega_c>1$, slow bath), which despite being both polar solvents have different characteristics. Methanol has a functional group \textit{OH} capable of releasing protons and the ability to form hydrogen bonds, while THF lacks these functional groups. Hence, the molecular systems interact much stronger with Methanol than with THF. For example, in the case of \textbf{\textit{BDP-Is}}, a decrease in the oscillations around $\sim 100$~fs in Methanol is observed (Fig.~\sugb{\ref{Fig5F}~(a)}), which has a $\tau_s=4$~ps ($\hbar\omega_c \sim10$~cm $^{-1}$), while registering a value of $\sim 160$~fs for the THF case (Fig.~\sugb{\ref{Fig5F}~(c)}) with $\tau_s = 1$~ps ($\hbar\omega_c\sim 40$~cm $^{-1}$).

The studied dyads exhibit  coherences at room temperature that are prolonged in the presence of THF (Fig.~\sugb{\ref{Fig5F}~(d)}), when compared to (Fig.~\sugb{\ref{Fig5F}~(b)}) for the same observables but in Methanol; this behavior is enhanced for the particular molecules \textbf{\textit{BDP-Is}} and \textbf{\textit{B2-Is}} for which the system-solvent  coupling strength $\eta$ follow the computed reorganisation energies. Furthermore, Fig.~\sugb{\ref{Fig5F}} allows the observation of the transition from coherent to incoherent oscillations, according to the  $\eta$ values: as $\eta$ increases, the energy exchange between the two-level electronic system and the bath becomes favored, and hence the two-level system oscillations amplitude are faster damped. This transition can be seen in both Fig.~\sugb{\ref{Fig5F}~(a)} and Fig.~\sugb{\ref{Fig5F}~(c)}. While for  $\eta=1.0193$, $\langle\sigma_z(t)\rangle$ exhibits damped coherent oscillations (Fig.~\sugb{\ref{Fig5F}~(a)}), the bath turns off all coherent features for a stronger coupling at  $\eta=3.0950$. In  Fig.~\sugb{\ref{Fig5F}~(c)}, the couplings are smaller and the transition from coherent to incoherent oscillations is not as sharp as in  Methanol. By comparing the effects due to Methanol (Fig.~\sugb{\ref{Fig5F}~(a)}, \sugb{(b)}) and THF (Fig.~\sugb{\ref{Fig5F}~(c)}, \sugb{(d)}) on the molecular electron dynamics, the population inversion oscillation amplitudes goes around  $\sim 0.94-1.0$ for the former, and about $\sim 0.88-1.0$ for the latter. As for the duration of such oscillations, in THF they last between $ \sim 50 $~fs and $\sim 180$~fs, while for Methanol such times are reduced to something between $40$~fs and $120$~fs which, again, evidences the stronger molecule-solvent coupling for the latter solvent.

\begin{figure*}[ht]
\centering
\includegraphics[scale=0.6]{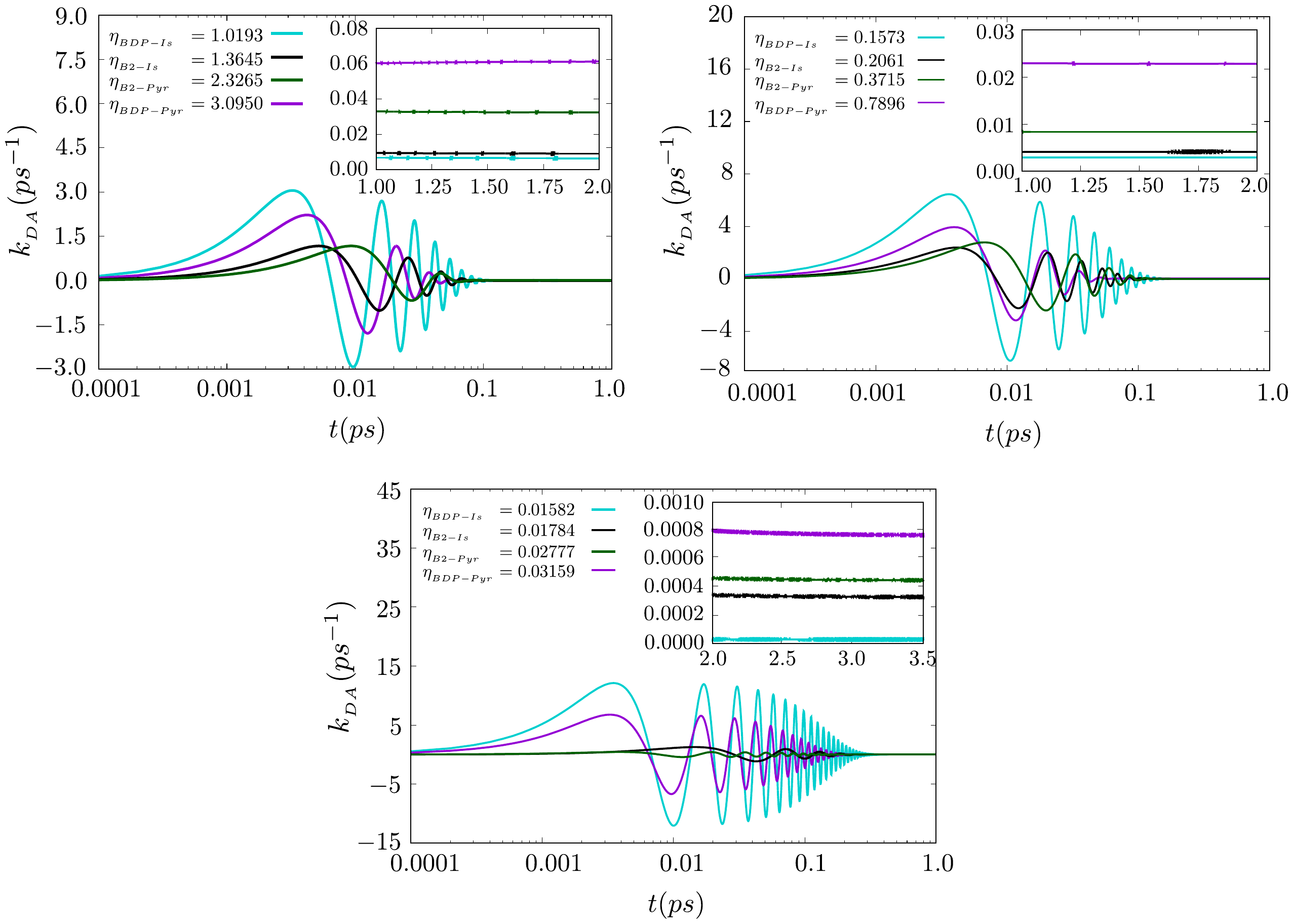}
\caption{Electron transfer rate $k_{_{DA}}$ for \textbf{\textit{BDP-Pyr}}, \textbf{\textit{BDP-Is}}, \textbf{\textit{B2-Pyr}} and \textbf{\textit{B2-Is}}, embedded in solvents: {(a)} Methanol, {(b)} THF, and  {(c)} Toluene. The boxes in the upper panels of each plot represent $k_{_{DA}}$ asymptotic behavior. The ultrafast electron population dynamics have been computed by using the HEOM method.}
\label{Fig7F}
\end{figure*}

\section{Electron Transfer Rates for the Two-Level System Configuration}

The electron transfer dynamics is, in general, more involved than a simple exponential decay. However, for sufficiently high temperatures and/or strong system-solvent couplings, the electronic population dynamics exhibit an exponential relaxation towards the equilibrium state.  In this case, the electronic population dynamics can be described by the simple kinetic equation:
\begin{eqnarray}
\dot{P}_D(t)& = -k_fP_D(t)+k_bP_A(t)\\
\dot{P}_A(t)& = -k_bP_A(t)+k_fP_D(t),\nonumber
\end{eqnarray}
where $k_f$ and $k_b$ denote the transfer rate from the donor to the acceptor and vice versa, respectively, and $P_D+P_A=1$. The stationary charge transfer rates are obtained as the limit when $t$ tends to infinity ($t\longrightarrow \infty$) of the time-dependent decay rate:
\begin{equation}
k_f=\lim_{t\longrightarrow \infty}k_f(t),
\end{equation}
with
\begin{equation}
k_f(t)=k_{_{DA}}(t)=-\frac{1}{2}\frac{\dot{P_D}(t)}{P_D(t)}.
\label{EcuF}
\end{equation}
Fig. ~\sugb{\ref{Fig7F}} shows the dynamics of the transfer rates according to Eq.~\sugb{\ref{EcuF}} in the presence of Methanol, THF and Toluene for the molecules described in Fig.~\sugb{\ref{Fig1F}}.
The asymptotic $t \longrightarrow \infty$ values  for the systems' transfer rates are shown in the upper panels, from which  higher   $k_{_{DA}}$ rates are observed for \textbf{\textit{BDP-Pyr}} and \textbf{\textit{B2-Pyr}} in all the solvents (Methanol, THF and Toluene). This result shows that the highest $k_{_{DA}}$ rates are found in molecular systems with a pyrrolidine ring, that is, the effect caused by the isozaxylin ring to increase the conjugation of fullerene does not favor intramolecular charge transfer rate. Furthermore, the incorporation of the alkoxyphenylethynyl groups, as in the \textbf{\textit{B2-Is}} and \textbf{\textit{BDP-Is}} systems, significantly reduce the asymptotic $k_{_{DA}}$ values, as can be seen in Fig.~\sugb{\ref{Fig7F}}. These results are tuned with the observation that the stronger the molecule-solvent coupling  $\eta$, the higher the asymptotic $k_{_{DA}}$ values (see Fig.~\sugb{\ref{Fig7F}}). 

For the case of the non-polar solvent here considered, Toluene, such $k_{_{DA}}$ values get significantly reduced, by two orders of magnitude in comparison to the polar solvents, Methanol and THF. In this sense, $k_{_{DA}}$ values are favored by polar environments. By  comparing the results for Methanol and THF, a higher $k_{_{DA}}$ is observed for the former (Fig.~\sugb{\ref{Fig7F}~(a)}) than for  the aprotic polar solvent THF; this may be due to the fact that Methanol has a functional hydroxylic group ($OH$) capable of yielding proton and has the ability to form hydrogen bridges that favor the  interaction with the considered molecular systems, in contrast to THF, that lacks these functional groups.

Finally, we would like to stress that the substitution of a nonpolar solvent with a polar one increases $k_{_{DA}}$ by one to two orders of magnitude. This also highlights the important result that the stronger the molecule-solvent coupling  $\eta$, the higher the asymptotic electron transfer rates $k_{_{DA}}$, regardless the coherence properties exhibited by the molecules (note that the damped coherence dynamics are maintained for a longer time in Toluene ($\sim 300$~fs), than in Methanol and THF ($\sim 100$~fs)).


\section{Effects due to the inclusion of a Zn-Porphyrin (ZnP) fragment as a molecular bridge}

 Photosensitizers play a critical role in light harvesting, charge transfer, and cell performance, and, in particular, in applications related to energy conversion device implementation~\sugb{\cite{ladomenou2014,gabrielsson2013}}. In this direction, porphyrin-based systems have attracted great attention, since they have a light collection capacity, a wide absorption consisting of intense bands ($400-450$~nm), and moderately intense $Q$ bands ($500-650$~ nm)~\sugb{\cite{jia2015, higashino2015}}, and photophysical and electrochemical properties that can be easily adjusted due to the existence of multiple meso- and $\beta$-modification sites. Furthermore, efficiencies of 10\% and 13\% through porphyrin dyes \textbf{YD-2} and \textbf{SM315}, respectively, have been achieved~\sugb{\cite{mathew2014, bessho2010}}; these adopt a push-pull type structure that has been shown to facilitate electron injection due to electron transfer from the donor site to the acceptor site, while the expansion of the $\pi$-conjugated linker and consideration of asymmetrical structures have been found to be beneficial for broadening and red-shifting the absorption of porphyrin compounds~\sugb{\cite{zhang2014, wang2014}}.

\begin{figure}[ht]
\centering
\includegraphics[scale=0.15]{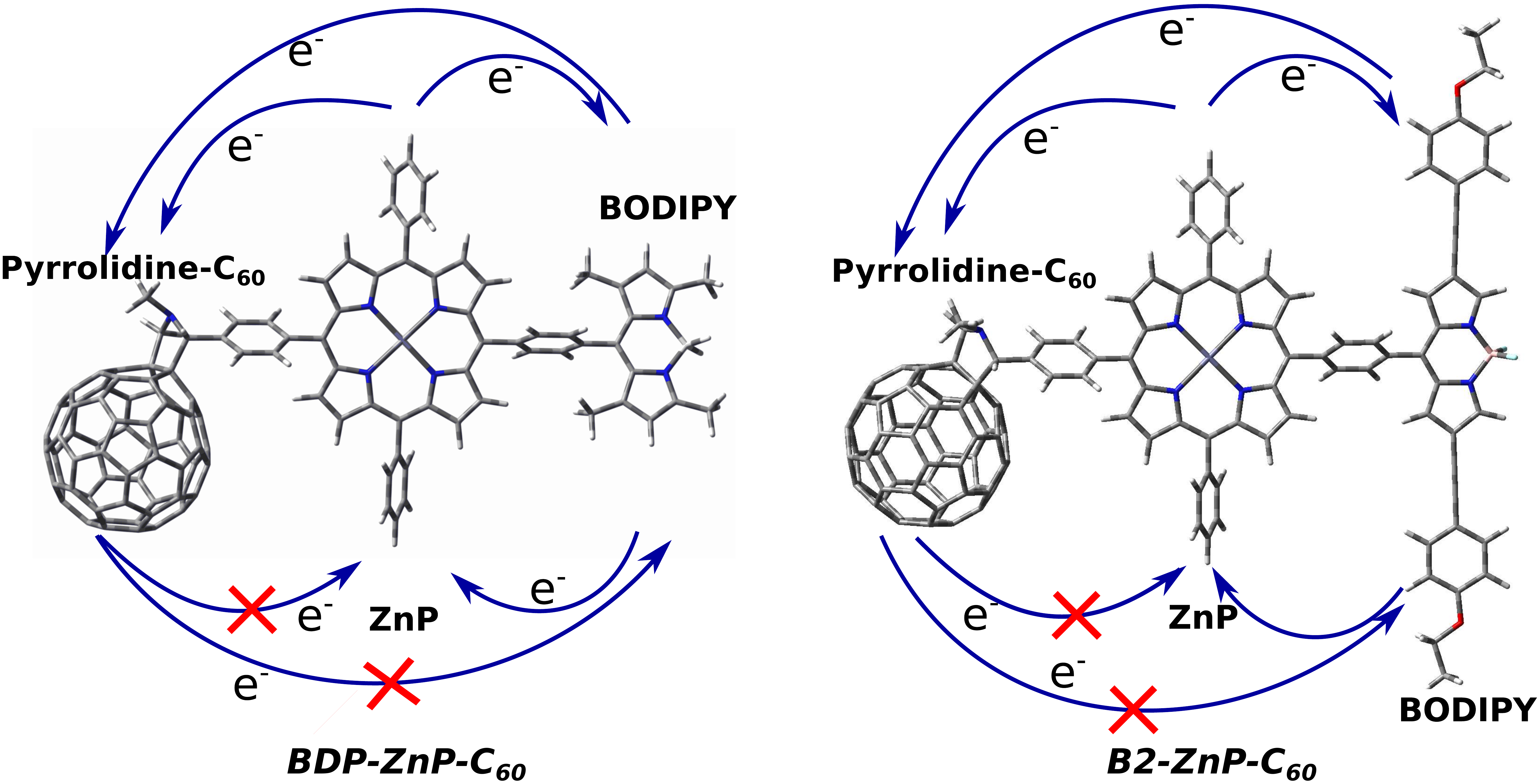} 
\caption{Structure and fragmentation scheme of the \textbf{\textit{BDP-ZnP-Fulleropyrrolidine}} (\textbf{\textit{BDP-ZnP-[PyrC$_{60}$]}}),  left, and \textbf{\textit{B2-ZnP-Fulleropyrrolidine}} (\textbf{\textit{B2-ZnP-[PyrC$_{60}$]}})~triad, right; electron transfer reactions in the triad are also shown.} 
\label{Fig8F}
\end{figure}

We next report on the  study of photoinduced charge separation in the fullerene \textbf{\textit{BDP-ZnP-[PyrC$_{60}$]}}~(\textbf{\textit{BDP-ZnP-C$_{60}$}}), and \textbf{\textit{B2-ZnP-[PyrC$_{60}$]}}~(\textbf{\textit{B2-ZnP-C$_{60}$}}) triad (see Fig.~\sugb{\ref{Fig8F}}), by means of  time-dependent DFT and the continuous model of polarization (C-PCM) for treating the environmental effects. The stabilization energy of charge transfer states by a polar medium depends significantly on whether the BODIPY fragment acts as an electron donor or as an electron acceptor.

In the considered compounds, electron-acceptor Fulleropyrrolidine and BODIPY (\textbf{\textit{BDP}} or \textit{\textbf{B2}}) are located at opposite sides of \textbf{\textit{[PyrC$_{60}$]}}, enabling the formation of six (6) different types of $CT$ states. The most likely $CT$ states, \textbf{\textit{B$^-$-ZnP$^+$-C$_{60}$}} ($CT_1$), and \textbf{\textit{B-ZnP$^+$-C$_{60}^-$}} ($CT_2$), are generated where \textbf{\textit{[PyrC$_{60}$]}} acts as an electron-donor. The other two,  \textbf{\textit{B$^+$-ZnP-C$_{60}^-$}} ($CT_3$), and \textbf{\textit{B$^+$-ZnP$^-$-C$_{60}$}} ($CT_4$), where \textbf{\textit{[PyrC$_{60}$]}} is an electron-acceptor. The last two types of $CT$ states are formed when charge separation occurs between the BODIPY and \textbf{\textit{C$_{60}$}} moieties (\textbf{\textit{B$^-$-ZnP-C$_{60}^+$}} and \textbf{\textit{B$^+$-ZnP-C$_{60}^-$}}). Initially, this study does not consider the states \textbf{\textit{B-ZnP$^-$ C$_{60}^+$}} and \textbf{\textit{B$^-$-ZnP-C$_{60}^+$}} because the ET is not present from \textbf{\textit{C$_{60}$}}, as shown in Fig.
~\sugb{\ref{Fig8F}}. 

\begin{table}[htp] 
\begin{center}
\centering
\caption{Parameters including the $ZnP$ fragment. Single excitation $E_x$ (eV),  oscillator strengths~$f_{osc}$, and weights of major Kohn-Sham (KS).}
\scalebox{0.78}{
\begin{tabular}{cccccccccc} 
\hline
\hline
\\
& & \multicolumn{2}{c}{\textbf{\textit{B2-ZnP-}[\textit{PyrC}$_{60}$]}} &  & \multicolumn{2}{c}{\textbf{\textit{B2-ZnP-}[\textit{PyrC}$_{60}$]}} & \\
\\
\hline
\\
 & \textbf{\textit{E}$_x$}  & \textbf{\textit{f}$_{osc}$} & \textbf{\textit{KS}} & \textbf{\textit{E}$_x$}  & \textbf{\textit{f}$_{osc}$} & \textbf{\textit{KS}} \\
\\
\hline
\hline
\\
\hspace{0.2 cm} $LE_1$  & 2.2037 & 0.6059 & H-1 $\rightarrow$ L   (0.99) &  2.9161 & 0.3573 & H-2 $\rightarrow$ L+3 (0.89)\\ 
\hspace{0.2 cm} $LE_2$  & 2.2488 & 0.0376 & H $\rightarrow$ L+5   (0.65) &  3.0561 & 0.0995 & H-1 $\rightarrow$ L+5 (0.73)\\
\hspace{0.2 cm} $CT_1$  & 2.1999 & 0.0013 & H $\rightarrow$ L     (0.99) &  2.4576 & 0.0004 & H-1 $\rightarrow$ L+2 (0.97)\\
\hspace{0.2 cm} $CT_2$  & 2.1849 & 0.0003 & H $\rightarrow$ L+1   (0.99) &  2.6991 & 0.0010 & H-2 $\rightarrow$ L+2 (0.99)\\
\hspace{0.2 cm} $CT_3$  & 1.9884 & 0.0001 & H-1 $\rightarrow$ L+1 (1.00) &  2.5889 & 0.0004 & H-1 $\rightarrow$ L+3 (0.98)\\
\hspace{0.2 cm} $CT_4$  & 2.7083 & 0.0081 & H-1 $\rightarrow$ L+4 (1.00) &  2.8752 & 0.0051 & H-2 $\rightarrow$ L+5 (0.99)\\
\\
\hline
\hline
\end{tabular}}
\label{Tab4_3}
\end{center}
\end{table}

\begin{figure*}[ht]
\centering
\includegraphics[scale=0.2]{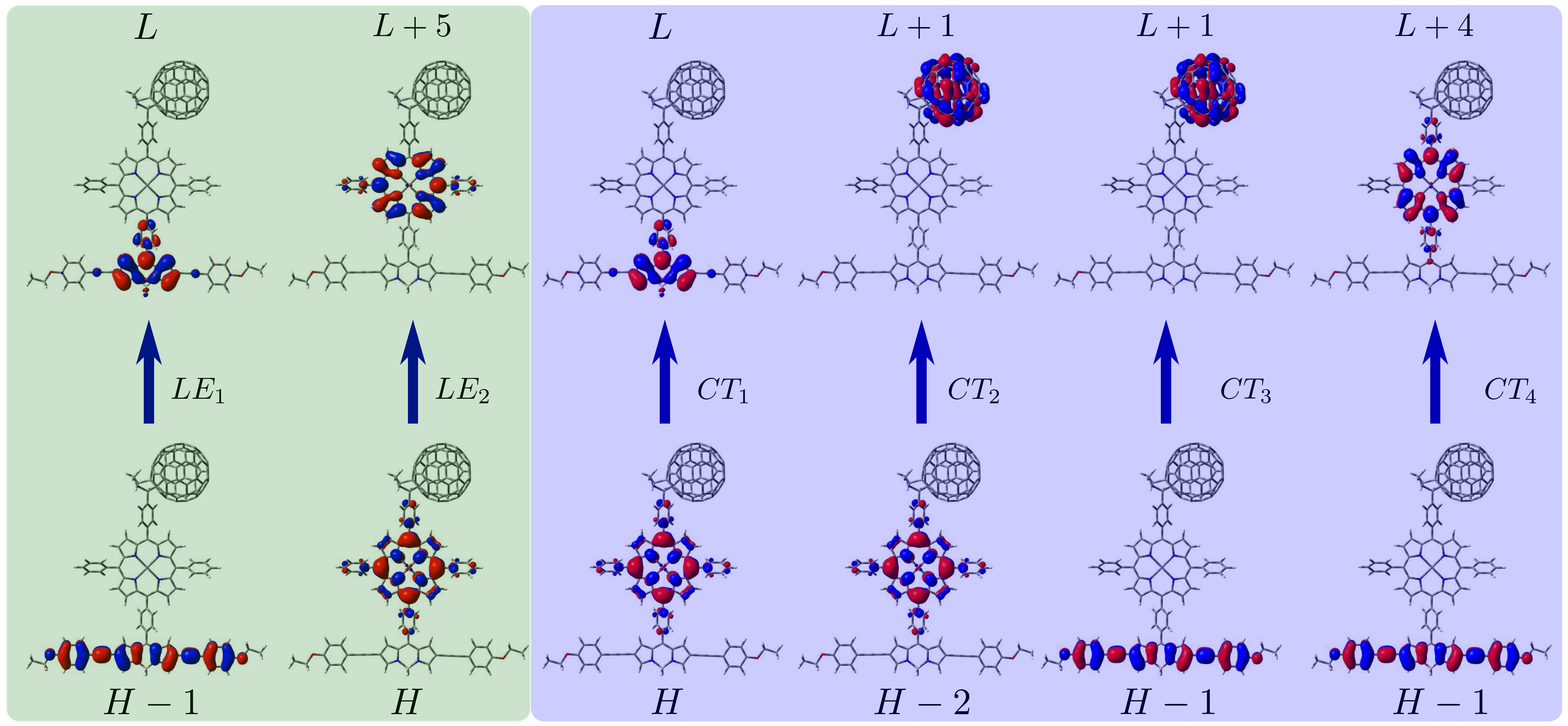}
\caption{Natural transition molecular orbitals representing the lowest-lying single $LE$ and $CT$ states for the \textbf{\textit{B2-ZnP-[PyrC$_{60}$]}}.}
\label{Fig10F}
\end{figure*}

The analysis of the excited states was carried out in terms of the exciton delocalization and charge transfer contributions. Two types of excited states can be identified: locally excited ($LE$) states, where an exciton is mostly localized on a single fragment, and $CT$ states.  The $LE_1$ state is characterized by a significant probability of light absorption (oscillator strength $f_{osc}=0.6059$), while the oscillator strength of the  $LE_2$ state is weaker for by an order of magnitude, $f_{osc}=0.0376$, for \textbf{\textit{B-ZnP-[PyrC$_{60}$]}}, as can be seen in Table~\sugb{\ref{Tab4_3}}. The NTOs for the two lowest $LE$ states, and four lowest $CT$ states are illustrated in Fig.~\sugb{\ref{Fig10F}}. As seen, the $LE_1$ and $LE_2$ excitations are almost completely localized on the BODIPY and the \textbf{\textit{ZnP}} fragment, whereas the $CT_1$ and $CT_2$ states are associated with ET from \textbf{\textit{ZnP}} to BODIPY and \textbf{C}$_{60}$, respectively. The $CT_3$ and $CT_4$ states are associated to ET from BODIPY to \textbf{C}$_{60}$ and \textbf{\textit{ZnP}}, respectively.

\begin{figure}[ht]
\centering
\includegraphics[scale=0.2]{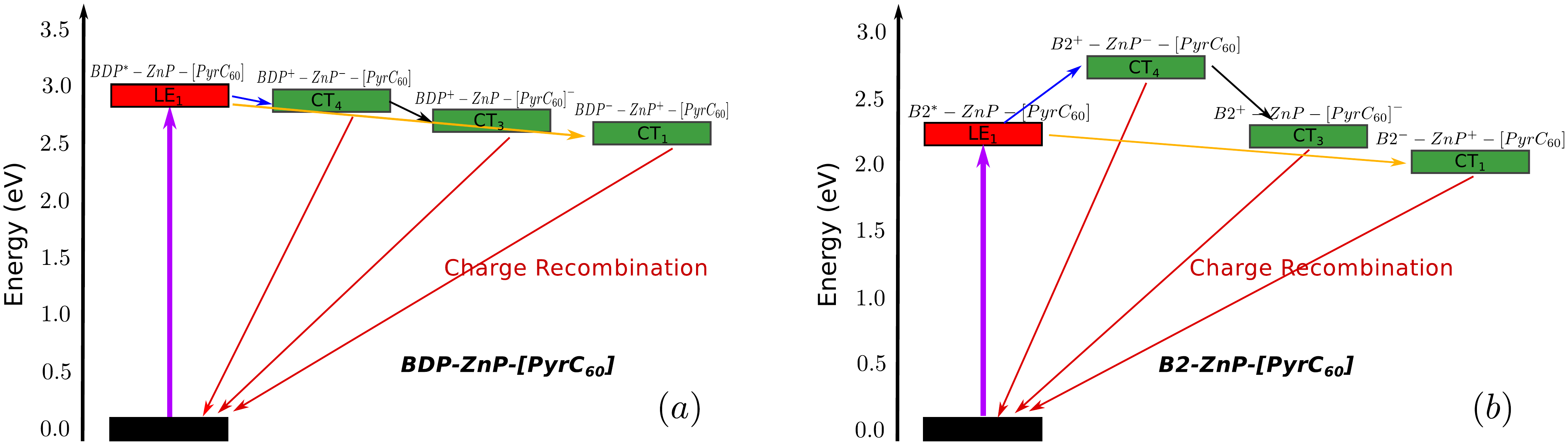}
\caption{Diagram of the reaction pathways from the initial locally excited state to the charge transfer states formed in the dyad molecules (a) \textbf{\textit{BDP-ZnP-PyrC$_{60}$}}, and (b) \textbf{\textit{B2-ZnP-[PyrC$_{60}$]}}.} 
\label{Fig9F}
\end{figure}

For molecular compounds \textbf{\textit{B2-ZnP-[PyrC$_{60}$]}} and \textbf{\textit{BDP-ZnP-[PyrC$_{60}$]}}, the process of electron transfer begins with the absorption of light by the electron-donor fragment BODIPY, this is followed by ET from BODIPY (\textit{\textbf{BDP}} or \textit{\textbf{B2}} for the second system) to \textbf{\textit{ZnP}} and \textbf{\textit{ZnP}} to \textbf{\textit{[PyrC$_{60}$]}}, to reach the target charge separation state (\textbf{\textit{B$^+$-ZnP-[PyrC$_{60}$]$^-$}}). The reaction pathways from the locally excited initial state (for $LE_1$) are illustrated in Fig.~\sugb{\ref{Fig9F}}. In addition to these electron transfer reactions, the systems can undergo recombination processes from any of the partially or fully separated states (red arrows). However, in this case we will not consider charge recombination processes towards the basis state.

\begin{figure*}[ht]
\centering
\includegraphics[scale=0.3]{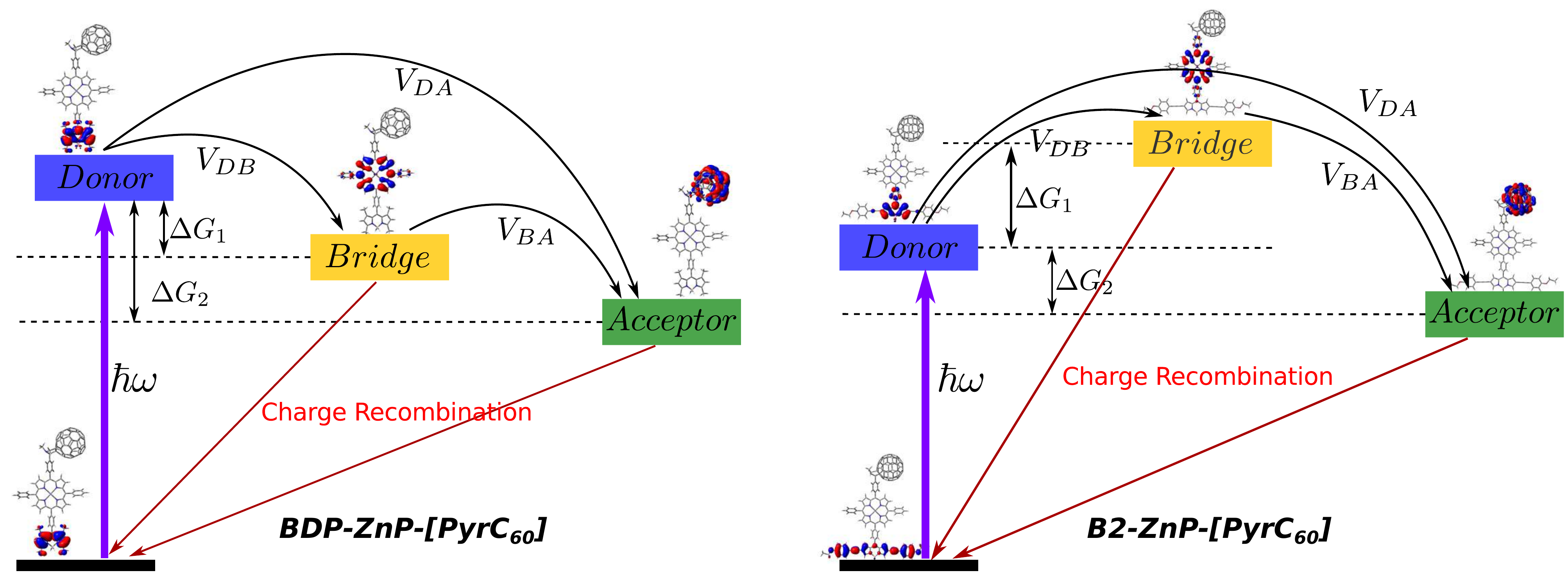}
\caption{Schematic image of a three-site $D$-$B$-$A$ system. The excitation energy of site $i=D,B,A$ is denoted $\varepsilon_i$ and the coupling between sites $i$ and $j$ by $V_{ij}$. The arrows symbolize the possible sequences that can be followed by the  transferred charge from the electron-donor  to the electron-acceptor fragments, in Methanol, for:  (a) \textbf{\textit{BDP-ZnP-[PyrC$_{60}$]}}, $\Delta G_1=-300$~cm$^{-1}$, $\Delta G_2=-1750$~cm$^{-1}$, $V_{DB}=100$~cm$^{-1}$, and $V_{BA}=760$~cm$^{-1}$, and (b) \textbf{\textit{B2-ZnP-[PyrC$_{60}$]}}, $\Delta G_1=4070$~cm$^{-1}$, $\Delta G_2=-150$~cm$^{-1}$, $V_{DB}=954$~cm$^{-1}$, and $V_{BA}=522$~cm$^{-1}$.}
\label{Fig11F}
\end{figure*}

The chosen method for calculating the electronic couplings is crucial. We use the GMH method to compute the coupling between the two diabatic states including many electron effects. In Fig.~\sugb{\ref{Fig11F}}, a scheme showing the calculated couplings is presented. Since the GMH method only couples two orbitals at a time, the coupling between the excited 
state \textbf{\textit{B$^*$-ZnP-[PyrC$_{60}$]}} and one charge transfer state, e.g., \textbf{\textit{B$^+$-ZnP$^-$-[PyrC$_{60}$]}} is the same as that between the other charge transfer state, \textbf{\textit{B$^-$-ZnP$^+$-[PyrC$_{60}$]}} and the charge separated state \textbf{\textit{B$^+$-ZnP-[PyrC$_{60}$]$^-$}}, as indicated with the  black and long orange lines in Fig.~\sugb{\ref{Fig9F}}. The GMH have three distinct recombination couplings (red lines).

The energies of the diabatic states are all given relative to the ground state. The energy differences between the partially charge separation states (CSSs), \textbf{\textit{B$^+$-ZnP$^-$-[PyrC$_{60}$]}} or \textbf{\textit{B$^+$-ZnP-[PyrC$_{60}$]$^-$}}, and the excited state \textbf{\textit{B$^*$-ZnP-[PyrC$_{60}$]}} are significantly different. For the molecular compounds \textbf{\textit{DBP-ZnP-[PyrC$_{60}$]}}, the state \textbf{\textit{B$^+$-ZnP$^-$-[PyrC$_{60}$]}} is 0.0408 eV (330 cm$^{-1}$) lower than the photoexcited state energy, but the state \textbf{\textit{B$^+$-ZnP-[PyrC$_{60}$]$^-$}} is 0.2169 eV (1750 cm$^{-1}$) lower than the referred photoexcited state energy. These states also show different couplings to the excited state, with GMH$_2$ more than twice as large as GMH$_1$. However, for the triad \textbf{\textit{B2-ZnP-[PyrC$_{60}$]}}, the state \textbf{\textit{B$^+$-ZnP$^-$-[PyrC$_{60}$]}} is 0.5046 eV (4070 cm$^{-1}$) higher than the photoexcited state energy, and the state \textbf{\textit{B$^+$-ZnP-[PyrC$_{60}$]$^-$}} is 0.0188 eV (150 cm$^{-1}$) lower  than that of such photoexcited state. 

Here we are not interested in studying transfer processes from the \textbf{\textit{ZnP}} fragment to the BODIPY fragment (energy transfer), hence the inclusion of the state \textbf{\textit{BDP$^-$-ZnP$^+$-[PyrC$_{60}$]}} (or \textbf{\textit{B2$^-$-ZnP$^+$-[PyrC$_{60}$]}}) is not considered in the system Hamiltonian, and such system can be analyzed in terms of the diagram shown in Fig.~\sugb{\ref{Fig11F}}, in which the electronic transfer of the electron-donor fragment to the molecular bridge, and from the bridge to the electron-acceptor fragment are considered. The parameters $\Delta G_{ij}$, represent the energy difference between the electron-donor state and that associated with the molecular bridge and the electron-acceptor fragment, with donor-bridge  $V_{_{DB}}$ and bridge-acceptor $V_{_{BA}}$ electronic couplings.

Most of the $CT$ states in the triad are characterized by a very weak oscillator strength and therefore the probability of their direct population is very low. However, they can be generated due to their interaction with the lower $LE$ states. The $CT$ states then undergo charge recombination reactions recovering the ground state. Both \textbf{\textit{ZnP}} and BODIPY exhibit highly absorbent bands. When the excitation is located in BODIPY ($LE_1$), states $CT_1$, $CT_3$, and $CT_4$ can be formed, while the decay of $LE_2$ located in \textbf{\textit{ZnP}} leads to the population of $CT_1$, $CT_2$ and $CT_4$. Since $LE_1$ and $LE_2$ energies are close to each other, an excitation transfer can occur between \textbf{\textit{ZnP}} and BODIPY. The deactivation of $LE$ can proceed through two competing reactions: i) the transfer of electrons, with the formation of  $CT$ states, and ii) the energy transfer between BODIPY and \textbf{\textit{ZnP}}. Thus, $LE_1$ can decay through the competing processes of electron transfer (formation of $CT$ states) and excitation energy transfer to $LE_2$.

\section{Electron Transfer Rates for the Three-Level System Configuration}

Understanding ET at the molecular level is critical to creating devices that exploit the properties of ever-smaller material length scales~\sugb{\cite{maggio2013, maggio2014}}. The donor-bridge-acceptor ($D$-$B$-$A$) chain  is an example of a very useful model system by which the effect of molecular structural parameters can be highly controlled and studied. Several different modeling approaches have been used to describe the properties of these systems, such as ET rates and magnetic exchange couplings~\sugb{\cite{weiss2005, berlin2008}}. 

\begin{figure}[ht]
\centering
\includegraphics[scale=0.33]{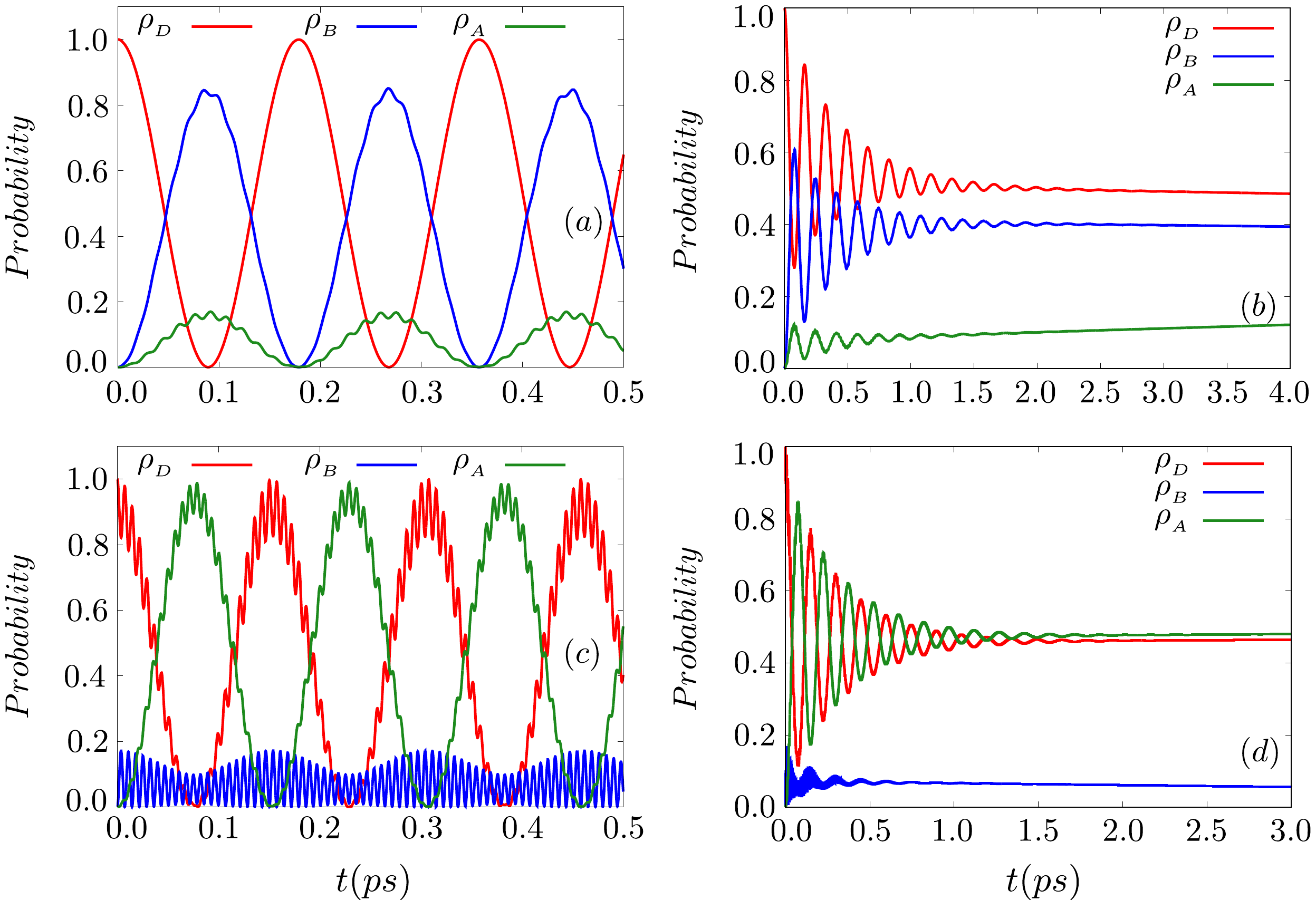} 
\caption{Dynamics of ultrafast electronic site populations for the donor ($D$),  bridge ($B$), and acceptor ($A$), $\rho_j=\langle\phi_j\vert\rho(t)\vert\phi_j\rangle$, $j=D,B,A$. 
The electronic-nuclear coupling is uniformly spread  over the bridge site. The model parameters are given in Table~\sugb{\ref{Tab4_3}}. (a)~and~(c) Coherent tunneling oscillations, $\eta=0$ (the fast oscillations are averaged for clarity) for the \textbf{\textit{BDP-ZnP-[PyrC$_{60}$]}} and \textbf{\textit{B2-ZnP-[PyrC$_{60}$]}} systems, respectively. Damped tunneling oscillations for the triads in Methanol for (b)  \textbf{\textit{BDP-ZnP-[PyrC$_{60}$]}}, $\eta = 1.8477$, and (d) \textbf{\textit{B2-ZnP-[PyrC$_{60}$]}}, $\eta = 2.4571$.} 
\label{Fig12F}
\end{figure}

The three-level system (3LS) represents the simplest possible $D$-$B$-$A$ molecular chain, parameterized by the three site energies and the two inter-site couplings, as shown in Fig.~\sugb{\ref{Fig11F}}. In some limiting cases of system parameters, the 3LS serves as an excellent simple model for analyzing possible jump mechanisms and superexchange electron transfer. In a jumping ET mechanism, the electrons population moves sequentially along a molecular chain from the donor site to the acceptor site, while in a superexchange mechanism the population tunnels from the donor site to the acceptor without significantly residing in any intermediate state (molecular bridge) and without direct donor-acceptor coupling~\sugb{\cite{renaud2013,paulson2005,berlin2002}}.

In order to describe the  three  level system
depicted in Fig.~\sugb{\ref{Fig11F}},  
 we first set $\varepsilon_{_D}=0$ to simplify the equations.
The 3LS Hamiltonian can then be written as~\sugb{\cite{may2008}}
\begin{equation}
H_{3LS}=\sum_{i=D,B,A}\varepsilon_i\vert\phi_i\rangle\langle\phi_i\vert + \sum_{i\neq j=D,B,A} V_{ij}\vert\phi_i\rangle\langle\phi_j\vert ,
\end{equation}
where $i$ is an index running over the three sites, and $\varepsilon_{_D} \neq \varepsilon_{_A} $. The states $\vert\phi_i\rangle$ are the $D$-$B$-$A$ site basis. In matrix representation, the 3LS in the $D$-$B$-$A$ basis reads:
\begin{equation*}
\widehat{H}_{3LS} = 
\begin{pmatrix}
   0   &     V_{_{DB}}    &       0      \\
V_{_{DB}} & \varepsilon_{_B} &     V_{_{BA}}    \\
   0   &     V_{_{BA} }   & \varepsilon_{_A}
\end{pmatrix},
\end{equation*}
where  $\varepsilon_{_B}$ and $\varepsilon_{_A}$ are the energies of the molecular bridge and the acceptor, respectively, and $V_{_{DB}}=V_{_{BD}}$.

\begin{figure*}[ht]
\centering
\includegraphics[scale=0.5]{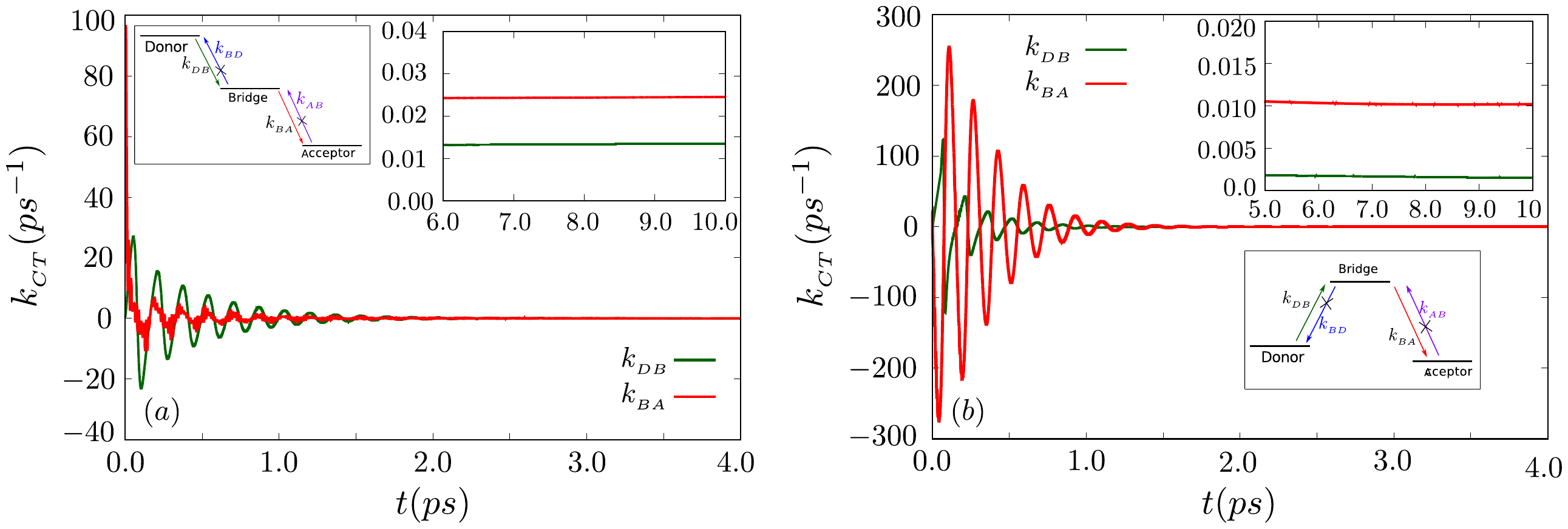} 
\caption{Dynamics of donor-bridge ($k_{DB}$) and bridge-acceptor ($k_{BA}$) charge transfer rates, calculated according to Eq.\sugb{12}. The fluctuations exhibited by the rates converge to the mean values indicated in the boxes, for (a) \textbf{\textit{BDP-ZnP-[PyrC$_{60}$]}}, and (b) \textbf{\textit{B2-ZnP-[PyrC$_{60}$]}}. The three-state kinetic schemes used are also indicated.} 
\label{Fig13F}
\end{figure*}

Figure~\sugb{\ref{Fig12F}} presents the dynamics of the electronic transfer after the instantaneous excitation of \textbf{\textit{B2-ZnP-[PyrC$_{60}$]}} or \textbf{\textit{BDP-ZnP-[PyrC$_{60}$]}}, in the absence ((a) and (c)) and presence ((b) and (d)) of Methanol. As expected from the studied conformation, an excitation initially located in the fragment \textbf{\textit{B2}} or \textbf{\textit{DBP}} is transferred and shared with \textbf{\textit{ZnP}} and \textbf{\textit{[PyrC$_{60}$]}} (see Fig.~\sugb{\ref{Fig9F}}). In the absence of nuclear-electronic coupling ($\eta=0$), the  $D$-$B$-$A$ model corresponds to a purely electronic dynamics, as would be observed in a completely rigid molecular system. Considering a high energy barrier $E_B-E_0>V_{DB}, V_{BA}$, the dynamics that follow an initial electronic preparation at the donor site is within the tunneling (superexchange) regime. Therefore, without considering relaxation mechanisms due to the bath, the electron exhibits an  ultrafast `jump' dynamics, periodically exchanging populations between the donor, and the bridge and acceptor sites. For example, it takes about  $t\sim 0.09$~ps for the donor site to depopulate and to transfer such population into $B$ and $A$ (Fig.~\sugb{\ref{Fig12F}(a)}), and this process becomes periodic in both cases (\textbf{\textit{BDP-ZnP-[PyrC$_{60}$]}} and \textbf{\textit{B2-ZnP-[PyrC$_{60}$]}}, Figs.~\sugb{\ref{Fig12F}(a)} and~\sugb{(c)}, respectively). Then the reverse jump process occurs, and a full cycle takes $1$ time unit; at time $t\sim 0.18$~ps a full period has taken place and again the population $\rho_D=1$. This said, in Fig.~\sugb{\ref{Fig12F}(a)} the bridge reaches much higher populations than the acceptor site, and the electron has the highest population density in the bridge when $\rho_D=0$.

On the other hand, when the  $D$-$B$-$A$ triad has a reflection $D$-$A$ symmetry, pronounced coherent oscillations of the electronic site population are observed between the donor and the recipient, as demonstrated in Fig.~\sugb{\ref{Fig12F}(c)}, in which three full periods of maximal transfer between the donor and the acceptor sites, with a total population oscillation period $t\sim 0.16$~ps, takes place.

Figure~\sugb{\ref{Fig12F}~(b)} and~\sugb{(d)} represent the electronic dynamics when the effects due to the environment are considered. Here we have, respectively, a reorganization energy $\lambda=30$~cm$^{-1}$ ($\eta=1.8477)$ for \textbf{\textit{BDP-ZnP-[PyrC$_{60}$]}}, and $\lambda=40$~cm$^{-1}$ ($\eta=2.4571)$ for \textbf{\textit{B2-ZnP-[PyrC$_{60}$]}}. Due to the interaction with the nuclear bath, now we have damped population oscillations in the electronic dynamics. The increased strength of the nuclear-electronic coupling leads to stronger (and faster) relaxation, eliminating coherent oscillations in the system. In Figs.~\sugb{\ref{Fig12F}~(b)} and~\sugb{(d)}, despite the strong coupling regime and ambient temperature considered, we find long-lasting oscillations for the donor and acceptor sites, beyond 1~ps.

Previous definitions of ``the rate of transport'' in the kinetic transport regime referred to a constant effective rate that dominates the disintegration of the donor to the recipient in a  $D$-$B$-$A$ biased  (asymmetric) triad~\sugb{\cite{felts1995}} or alternatively to a steady-state current in the presence of a source and drain terms~\sugb{\cite{segal2000, weiss2006}}. For an electronically constrained $D$-$B$-$A$ system, neither of these definitions is consistent with the electron transfer kinetics, and there is no such single constant rate. Instead, modulo  a short transient time $\tau$ corresponding to a coherent electronic dynamics in the bridge (see Fig.~\sugb{\ref{Fig12F}~(b)} and \sugb{(d)}), the  diagonal elements of the electronic density matrix follow a kinetic scheme corresponding to three effective electronic sites and two effective speed constants, as outlined in the box in Figs.~\sugb{\ref{Fig13F}~(a)} and \sugb{(b)}. The constant rates  are defined by the following kinetic equations (for $t>\tau$)~\sugb{\cite{segal2000, nitzan2006, blumberger2015}}:
\begin{eqnarray}
\frac{dP_n}{dt} &=& -(k_{_{n-1\leftarrow n}} +k_{_{n+1\leftarrow n}})P_{_n}+k_{_{n\leftarrow n-1}}P_{_{n-1}}\nonumber
\\ && 
+k_{_{n\leftarrow n+1}}P_{_{n+1}},
\end{eqnarray}
where $P_n$ is the population at site $n$, with  $n=D,B,A$. We then have:
\begin{equation}
\begin{pmatrix}
\dot{P}_{_D}(t) \\
\dot{P}_{_B}(t) \\
\dot{P}_{_A}(t) 
\end{pmatrix}
=\begin{bmatrix}
 -k_{_{DB}} &       k_{_{BD}}     &  0      \\
  k_{_{DB}} & -(k_{_{BD}}+k_{_{BA}}) &  k_{_{AB}} \\
  0      &       k_{_{BA}}     & -k_{_{AB}} \\
\end{bmatrix}
\begin{pmatrix}
P_{_D}(t) \\
P_{_B}(t) \\
P_{_A}(t) 
\end{pmatrix},
\end{equation}
where $k_{_{DB}}$ corresponds to the transport from the donor to the bridge  and $k_{BA}$ to that from the bridge to the acceptor. The rates can be calculated `on the fly', by  means of the following equations
\begin{equation}
k_{_{DB}} = -\frac{\dot{P}_{_D}(t)}{P_{_D}(t)},\nonumber
\end{equation}
\begin{equation}
k_{_{BA}} =-\frac{\dot{P}_{_B}(t)+\dot{P}_{_D}(t)}{P_{_B}(t)},
\end{equation}
\label{Ecu_Rate}
where the reverse transfer rates $k_{_{BD}}$ and $k_{_{AB}}$ are not considered. When the kinetic scheme is valid, the calculated rates give almost constant values (as function of time), as demonstrated in the box in Figs.~\sugb{\ref {Fig13F} (a)} and \sugb{(b)}.

The rate of electron (or exciton) transfer is controlled by three parameters: the electronic coupling $V_{_{ij}}$ of the  involved states, the reorganization energy $\lambda$, and the Gibbs energy $\Delta G^0$. The reorganization energy is usually divided into two parts, $\lambda = \lambda_i + \lambda_s$. The $\lambda_i$ contribution is required to rearrange nuclei of the system, and the solvent term $\lambda_s$ is due to changes in solvent polarization. The internal reorganization energy was estimated by using  the energy differences of the anion- and cation-radicals taken in their equilibrium geometries as well as at geometries of the neutral species. Solvent reorganization energy was accounted for the entire excited states of interest using a polarizable continuum model (C-PCM) in the monopole approximation.

Comparison of the charge transfer rates in Fig.~\sugb{\ref{Fig13F}} shows that the inclusion of the alkoxyphenylmethyl group decreases the electronic transfer between the donor-bridge and the bridge-acceptor fragments. In addition, it considerably affects the energy levels and the conformational shape of the donor, bridge, and acceptor states, as seen in the upper left boxes and main ghraphs of Figs.~\sugb{\ref{Fig13F}(a)} and \sugb{(b)}. Here, a stepped type is observed for the  \textbf{\textit{BDP-ZnP-[PyrC$_{60}$]}} system and a $\Lambda$-type for the states of the system \textbf{\textit{B2-ZnP-[PyrC$_{60}$]}}. 

The results for both $k_{_{DB}}$ and $k_{_{BA}}$ show better and higher values for \textbf{\textit{BDP-ZnP-[PyrC$_{60}$]}} than for \textbf{\textit{B2-ZnP-[PyrC$_{60}$]}}, where the alkoxyphenylmethyl group has been included. We point out that the decrease in $k_{_{CT}}$ is related to the negative value of $\Delta G_{_{ij}}$, and hence the greater the value of $-\Delta G_{_{ij}}$ the greater $k_{_{CT}}$, as seen in Fig.~\sugb{\ref{Fig13F}}. For the \textbf{\textit{B2-ZnP-[PyrC$_{60}$]}} triad, we obtain  $k_{_{BA}}>k_{_{DB}}$, where $-\Delta G_{_{DB}}< -\Delta G_{_{DA}}$. The same order also  holds for the \textbf{\textit{BDP-ZnP-[PyrC$_{60}$]}} triad. However, the charge transfer rates $k_{_{CT}}$ for the latter triad are considerably higher compared to the former one, since the $-\Delta G_{_{DB}}$ values also increase considerably if compared to the former triad.

Finally, we stress that the higher transfer rate obtained for the $D$-$B$-$A$ systems corresponds to the  \textbf{\textit{BDP-ZnP-[PyrC$_{60}$]}} triad, which interestingly, and in contrast with the results for the  $D$-$A$ systems,  exhibits prolonged coherence and populations oscillations, beyond 1~ps.

Although good results have been obtained on charge transfer processes, it should be mentioned that the environment is described in terms of a continuous pattern where the solvent is treated as homogeneous dielectric. Therefore, it is possible to improve the results considering that at molecular scales the surrounding media are no longer continuous as shown by Mennucci et al.~\sugb{\cite{curutchet2011}}, Which has shown significant changes in transfer rates. In addition, one could think of considering the dielectric constant differently since charge transfer processes are very fast and a dielectric constant of this form would not be appropriate as recently demonstrated by Havenith et al.\sugb{\cite{sami2020}} However, the modeling of the environment (solvent) as a bath of independent harmonic oscillators has been widely used in the study of quantum dissipation, crucial to understanding the effects of quantum coherence, energy transfer, and charge transfer~\sugb{\cite{reina2002, Scholes2017, collini2009, chenu2015}}, present in the processes of photosynthesis, vibrational energy redistribution and other~\sugb{\cite {madrid2019, wittmann2020, romero2014, ishizaki2012}}

\section{Conclusions}

The dynamics of electron transfer of molecular systems built from BODIPY and Fullerene derivatives were simulated by using DFT, TD-DFT, and HEOM formalisms. These allowed to construct the full diabatic Hamiltonians of the molecular systems, here treated as open quantum systems, by considering the effects of their electronic structure (pyrrolidine and isozaxylin rings) and the polarity of the different solvents to which they were coupled. 

We found that, for the case of donor-acceptor ($D$-$A$) systems, here described as TLSs, the electron transfer dynamics are very sensitive to solvent polarity, being the  asymptotic transfer rates favored for solvents with the highest dielectric constants $\epsilon_{_j}$ ($j$ stands for Methanol, THF, and Toluene), and larger system-solvent strengths coupling $\eta$, in  all the studied  molecular systems. In addition, with respect to the incorporation of the alkoxyphenylethynyl groups, in Methanol and THF solvents, a considerable $k_{_{CT}}$ decrease  was observed for the systems with an isoxazoline ring, while  systems with a pyrrolidine  ring have an increase in $k_{_{CT}}$; in both cases the $k_{_{CT}}$ rates fall within the same order of magnitude (and are smaller for THF than Methanol). However, in the presence of nonpolar Toluene, a considerable decrease in $k_{_{CT}}$, by two orders of magnitude,  takes place by the alkoxyphenylethynyl groups in both $Is$ and $Pyr$ rings. Regardless the electron coherence properties exhibited by the BODIPY and Fullerene derivatives-based compounds, the molecular charge transfer is enhanced by stronger system-solvent coupling, in the incoherent regime. 

In the case of the donor-bridge-acceptor ($D$-$B$-$A$) systems,  a Zn-Porphyrin molecular bridge was incorporated to generate the \textbf{\textit{BDP-ZnP-C$_{60}$}} and \textbf{\textit{B2-ZnP-C$_{60}$}} triads, for which we have computed their structure and  excited states properties within an effective 3LSs approach. Six types of $CT$ states have been identified. The triads demonstrate a markedly different stabilization of the $CT$ states for polar and nonpolar media, where BODIPY acts as an electron donor. Furthermore, the anionic and cationic BODIPY radicals have quite different solvation energies. This explains the marked sensitivity of  $CT_{3}$ (\textbf{\textit{B$^+$-ZnP-[PyrC$_{60}$]$^-$}}) and $CT_5$ (\textbf{\textit{B$^-$-ZnP-[PyrC60]$^+$}}) to the solvent polarity. The analysis of the calculated ET rates shows that additional deactivation channels from the porphyrin excited state may come into play as the solvent polarity increases. Our results predict a low coupling from the photoexcited \textbf{\textit{BDP$^*$-ZnP-C$_{60}$}} state to \textbf{\textit{BDP$^+$-ZnP$^-$-C$_{60}$}}, at the partial charge separation (CS) state with 0.041 eV lower in energy than the photoexcited state. The dynamics also show that this state remains largely unpopulated. Instead, we see an initial transfer to the \textbf{\textit{B2$^+$-ZnP$^-$-C$_{60}$}} state, which is 0.504 eV higher in energy than the photoexcited state. From this partial CSS, the full $CT$ state \textbf{\textit{B2$^+$-ZnP-C$_{60}$}} (or \textbf{\textit{BDP$^+$-ZnP-C$_{60}$}}) are populated, showing a stronger transfer process for the `cascade type' system.

Our results for the considered $D$-$B$-$A$ systems show, in contrast with that observed for the  $D$-$A$ molecular   dyads, that the largest charge transfer rates are obtained for triads with long lasting ($\sim $~1~ps) electron  coherence and populations oscillations. Hence,
the results here presented suggest that in molecular system designs that exhibit and/or take advantage of coherent electron transfer, maintaining close energy levels and large electron coupling values is critical. This optimal regime can be adjusted by an appropriate choice of  solvent and the electronic structure of molecular donors and acceptors.

\section*{Acknowledgements}

This work was supported by the Colombian Science, Technology and Innovation Fund-General Royalties System---``Fondo CTeI--Sistema General de Regal\'ias'' (contract No. BPIN 2013000100007).



\bibliography{Bibliografia}

\end{document}